\RequirePackage{ifpdf}
\ifpdf 
\documentclass[pdftex]{sigma}
\else
\documentclass{sigma}
\fi

\numberwithin{equation}{section}

\begin{document}

\allowdisplaybreaks
	
\renewcommand{\PaperNumber}{099}

\FirstPageHeading

\renewcommand{\thefootnote}{$\star$}

\ShortArticleName{From Ising Form Factors to $n$-Fold Integrals}

\ArticleName{From Holonomy
 of the Ising Model Form Factors to\\ $\boldsymbol{n}$-Fold Integrals
and the Theory of Elliptic Curves\footnote{This paper is a contribution to the Proceedings of
the Seventh International Conference ``Symmetry in Nonlinear
Mathematical Physics'' (June 24--30, 2007, Kyiv, Ukraine). The
full collection is available at
\href{http://www.emis.de/journals/SIGMA/symmetry2007.html}{http://www.emis.de/journals/SIGMA/symmetry2007.html}\\[2mm]
\textit{Dedicated to the memory of Vadim Kuznetsov.}}}
			
\Author{Salah BOUKRAA~$^\dag$, Saoud HASSANI~$^\ddag$,
Jean-Marie MAILLARD~$^\S$ and Nadjah ZENINE~$^\ddag$}

\AuthorNameForHeading{S.~Boukraa, S.~Hassani,
J.-M.~Maillard and N.~Zenine}

\Address{$^\dag$ LPTHIRM and D\'epartement d'A{\'e}ronautique,
Universit\'e de Blida, Algeria}
\EmailD{\href{mailto:boukraa@mail.univ-blida.dz}{boukraa@mail.univ-blida.dz}}

\Address{$^\ddag$ Centre de Recherche Nucl\'eaire d'Alger,
2 Bd. Frantz Fanon, BP 399, 16000 Alger, Algeria}
\EmailD{\href{mailto:okbi_h@yahoo.fr}{okbi\_h@yahoo.fr}, \href{mailto:njzenine@yahoo.com}{njzenine@yahoo.com}}

\Address{$^\S$ LPTMC, Universit\'e de Paris 6, Tour 24,
 4\`eme \'etage, case 121, 4 Place Jussieu,\\
 $\phantom{^\S}$~75252 Paris Cedex 05, France}
\EmailD{\href{mailto:maillard@lptmc.jussieu.fr}{maillard@lptmc.jussieu.fr}, \href{mailto:maillard@lptl.jussieu.fr}{maillard@lptl.jussieu.fr}}

\ArticleDates{Received September 19, 2007, in f\/inal form October 07, 2007; Published online October 15, 2007}

\Abstract{We recall the form factors $ f^{(j)}_{N,N}$
corresponding to the $ \lambda$-extension
 $ C(N,N; \lambda)$ of the two-point diagonal
 correlation function of the
Ising model on the square lattice and their associated linear
dif\/ferential equations which exhibit
 both a ``Russian-doll'' nesting,
and a~decomposition of the linear
dif\/ferential operators as a direct sum of operators (equivalent
 to symmetric powers of the dif\/ferential operator
of the complete elliptic integral $E$).
The scaling limit
of these dif\/ferential operators breaks the direct sum
structure but not the ``Russian doll'' structure, the ``scaled''
linear dif\/ferential operators
being no longer Fuchsian. We then introduce some multiple integrals
of the Ising class expected to have the same singularities as the
singularities of the $n$-particle contributions $\chi^{(n)}$
to the susceptibility of the square lattice Ising model.
We f\/ind the Fuchsian linear dif\/ferential equations satisf\/ied by these
 multiple integrals for $n=1, 2, 3, 4$ and,
only modulo a prime, for $n=5$ and $6$, thus providing a large set
of (possible) new singularities of the $\chi^{(n)}$.
We get the location of these singularities
by solving the Landau conditions.
We discuss the mathematical, as well as physical,
interpretation of these new singularities.
Among the singularities found,
we underline the fact that the quadratic polynomial condition
$ 1+3 w +4 w^2 = 0$, that occurs in
the linear dif\/ferential equation of $ \chi^{(3)}$,
actually corresponds to
 the occurrence of complex multiplication for
elliptic curves. The interpretation of complex multiplication
 for elliptic curves as complex f\/ixed points of
 generators of the exact renormalization group
is sketched. The other singularities occurring in our multiple integrals
are not related to complex multiplication situations, suggesting
a geometric interpretation in terms of more general
(motivic) mathematical structures
beyond the theory of elliptic curves.
The scaling limit of the (lattice of\/f-critical) structures
as a conf\/luent limit of regular singularities is discussed in the conclusion.}


\Classification{34M55; 47E05; 81Qxx; 32G34; 34Lxx; 34Mxx; 14Kxx}

\renewcommand{\thefootnote}{\arabic{footnote}}
\setcounter{footnote}{0}

\section{Introduction}
\label{recalls}

This paper displays a selection of works and results that
have been obtained by the authors in collaboration
with B.M.~McCoy, W.~Orrick and J.-A.~Weil. It also provides
new ideas and viewpoints at the end of Subsection~\ref{equiv2},
in Section~\ref{morebrid} and in the Conclusion~\ref{conclu}. We also give new
results of linear dif\/ferential operators modulo prime that had not
 been published before in Appendix~\ref{subsec} and Appendix~\ref{subsec6}.

The two dimensional Ising model in zero magnetic
 f\/ield is, historically, the most
important solvable model in all of theoretical physics. The free
energy~\cite{ons1}, the partition
function on the f\/inite lattice~\cite{kauf}
and the spontaneous magnetization~\cite{ons2,yang} were computed
long ago by Onsager, Kaufman and Yang. These computations, and
subsequent studies of the correlation
 functions~\cite{ko,wu-mc-tr-ba-76},
 form the basis of
scaling theory and of the renormalization group approach to critical
phenomena.

Let us f\/irst recall the form factors~\cite{Holo} of the lattice Ising model.
Our starting point will be the expansions of the {\em diagonal} correlations
in an exponential form~\cite{wu-mc-tr-ba-76}, both for $T<T_c$
\begin{gather}
\label{cm}
C_{-}(N,N) =
(1-t)^{1/4}
\exp \left(\sum_{n=1}^{\infty} F^{(2n)}_{N,N} \right)
\end{gather}
with
\begin{gather*}
t= \left( \sinh(2E^v/k_BT)\sinh(2E^h/k_BT) \right)^{-2}
\end{gather*}
and for $T>T_c$
\begin{gather}
\label{cp}
C_{+}(N,N) =
 (1-t)^{1/4} \sum_{n=0}^{\infty}
 G^{(2n+1)}_{N,N}  \exp
 \left( \sum_{n=1}^{\infty} F^{(2n)}_{N+1,N+1} \right)
\end{gather}
with
\begin{gather*}
t = \left( (\sinh(2E^v/k_BT)\sinh(2E^h/k_BT) \right)^2
\end{gather*}
where $E^h$ and $E^v$ are the horizontal and vertical interaction
energies of the Ising model. We will restrict in the following
to the isotropic Ising model. For diagonal correlation functions,
there is no dif\/ference between the isotropic and anisotropic
models: the diagonal correlations are functions of the modulus
$k = \sinh(2E^v/k_BT)\sinh(2E^h/k_BT)$.
The dif\/ference comes with of\/f-diagonal modes and is
sketched in~\cite{Fuchs}.

When the exponentials in (\ref{cm}) and (\ref{cp}) are expanded,
the correlations
 can also be written in what is called a ``form factor'' expansion
\begin{gather}
\label{formm}
C_{-}(N,N) =(1-t)^{1/4}
\left(1+\sum_{n=1}^{\infty} f^{(2n)}_{N,N}\right), \\
C_{+}(N,N) =
(1-t)^{1/4} \sum_{n=0}^{\infty} f^{(2n+1)}_{N,N}.
\label{formp}
\end{gather}
The form factor $f^{(j)}_{N,N}$ is interpreted as the ``$j$-particle''
contribution to the two-point correlation function.
 It is natural to consider
$\lambda$-extensions~\cite{mtw,wu-mc-tr-ba-76} of the previous functions
\begin{gather}
\label{formm1}
C_{-}(N,N;\lambda) = (1-t)^{1/4}
\left(1+\sum_{n=1}^{\infty}\lambda^{2n} f^{(2n)}_{N,N}\right), \\
C_{+}(N,N;\lambda) = (1-t)^{1/4}
\sum_{n=0}^{\infty} \lambda^{2n}  f^{(2n+1)}_{N,N} \label{formp1}
\end{gather}
which weight each $ f^{(j)}_{N,N}$ by some power
 of $ \lambda$, and to interpret $ \lambda$
as being analogous to a coupling constant in a quantum f\/ield theory expansion.
Such $ \lambda$-extensions naturally
emerge from the {\em Fredholm determinant
framework} in~\cite{wu-mc-tr-ba-76}.
We will present {\em new integral representations}
 for $ F^{(2n)}_{N,N}$, $G^{(2n+1)}_{N,N}$ and
 $ f^{(j)}_{N,N}$ in Section~\ref{new}. We
 will see that they are much simpler,
 and more transparent, than the forms obtained
 from $ C(M,N)$ of~\cite{wu-mc-tr-ba-76}
 by specializing to $ M= N$.

The diagonal correlations $ C(N,N)$ have the property, discovered by
 Jimbo and Miwa~\cite{jm} in 1980, that
their log-derivatives are solutions of
the ``sigma'' form\footnote{We use
a variable $ t$ which is the inverse of the one of
Jimbo and Miwa~\cite{jm}.}
 of a Painlev{\'e} VI equation
\begin{gather}
\left( t (t-1) {d^2\sigma \over dt^2} \right)^2
 = N^2  \left( (t-1) {d\sigma \over dt} -\sigma \right)^2
-4 {d\sigma \over dt}  \left((t-1) {d\sigma\over dt}
-\sigma -{1\over 4} \right)
 \left(t {d\sigma \over dt} -\sigma \right)\label{jimbo-miwa}
\end{gather}
where $ \sigma $ is def\/ined for $ T < T_c $ as
\begin{gather*}
\sigma_N(t) =
 t  (t-1){d \ln C_{-}(N,N)\over dt} -{t\over 4}
\end{gather*}
with the normalization condition
\begin{gather}
\label{cmnorm}
C_{-}(N,N) = 1 +O(t)
 \qquad {\rm for} \qquad  t\rightarrow 0
\end{gather}
and, for $ T > T_c$, as
\begin{gather*}
\sigma_N(t) =
t  (t-1) {d \ln C_{+}(N,N) \over dt} -{1\over 4}
\end{gather*}
with the normalization condition
\begin{gather}
\label{cpnorm}
C_{+}(N,N) =
{(1/2)_N\over N!}  t^{N/2}  \left(1+O(t)\right)
 \qquad {\rm for} \qquad  t\rightarrow 0,
\end{gather}
where $ (a)_N = \Gamma(a+N)/\Gamma(a) $
 denotes the Pochhammer symbol.

One can easily verify that (\ref{jimbo-miwa}), the $ N$-dependent
 sigma form of Painlev\'e VI, is actually
{\em covariant by the Kramers--Wannier duality}
\begin{gather*}
(t, \sigma, \sigma', \sigma'') \quad  \rightarrow \quad
\left( {{1} \over {t}}, {{\sigma} \over {t}},
 \sigma - t  \sigma',
 t^3  \sigma''
\right).
\end{gather*}

On another hand, Jimbo and Miwa introduced in~\cite{jm}
an {\em isomonodromic} $ \lambda$-extension of $ C(N,N)$.
Remarkably this more general function $ C(N,N; \lambda)$
also satisf\/ies~\cite{PainleveFuchs,or-ni-gu-pe-01b}
the Pain\-le\-v{\'e}~VI equation (\ref{jimbo-miwa}). The motivation of
introducing an isomonodromic parameter $ \lambda$,
in the framework of {\em isomonodromy deformations},
 is, at f\/irst sight, quite dif\/ferent
 from the ``coupling constant Fredholm-expansion''
 motivation at the origin
of the form factor $ \lambda$-extensions~(\ref{formm1}) and~(\ref{formp1}).
In~\cite{PainleveFuchs}
we have shown that these two $ \lambda$-extensions
 are {\em actually the same} by demonstrating that
the recursive solutions of (\ref{jimbo-miwa}),
analytic\footnote{The $ \lambda$-extensions (\ref{formm1}) and (\ref{formp1}) are
analytic at $ t \sim 0$
in $ t$ for $ T < T_c$ and, when
 $ T > T_c$, analytic in $ t$ for $ N$ even, and
in $ t^{1/2}$ for $ N$ odd.} in $ t^{1/2}$,
agree with (\ref{formm1}) and
(\ref{formp1}) where the $ f^{(j)}_{N,N}$'s are
 obtained from $ C_{\pm}(N,N; \lambda)$,
 the $ \lambda$-extension of $ C_{\pm}(N,N)$.
The normalization condition (\ref{cmnorm}) f\/ixes
 one integration constant in the solution to (\ref{jimbo-miwa}).
 We f\/ind that the second integration constant is
 a free parameter, and, denoting that parameter
 by $ \lambda$, we f\/ind that our one-parameter family
of solutions for $ C_{-}(N,N)$ can be written
 in a form structurally similar to the
right hand side of (\ref{formm1}). Furthermore, we have conf\/irmed,
 by comparison with series expansions of the multiple
integral formulas for $ f_{N,N}^{(j)}$
 derived in Section~\ref{new}, that this family of
 solutions is, in fact, identical to $ C_{-}(N,N;\lambda)$
 as def\/ined in (\ref{formm1}). Similarly, the
 condition (\ref{cpnorm}) gives rise to a~one-parameter
 family of solutions for $ C_{+}(N,N) $ that is
identical to (\ref{formp1}).

\section[New integral representations for the $ f^{(n)}_{N,N}$'s]{New integral representations for the $\boldsymbol{f^{(n)}_{N,N}}$'s}\label{new}

The form factor expressions for the two-point
correlation functions $ C(M,N)$
of~\cite{nickel1, nickel2, pt,or-ni-gu-pe-01b,wu-mc-tr-ba-76,yamada}
are obtained by expanding the exponentials in (\ref{cm}), and (\ref{cp}),
in the form given in~\cite{wu-mc-tr-ba-76} as multiple integrals,
and integrating over half the variables. The form of the result depends
on whether the even, or odd, variables of~\cite{wu-mc-tr-ba-76} are integrated
out. For the general anisotropic lattice, one form of this result is
given, for arbitrary $ M$ and $ N$, in~\cite{or-ni-gu-pe-01b}.
When specialized to the isotropic case the result is
\begin{gather*}
 f^{(2j)}_{M,N} = \hat{C}^{2j}(M, N), \qquad
 f^{(2j+1)}_{M,N} = {{\hat{C}^{2j+1}(M, N)} \over {s}},
\end{gather*}
where $ s$ denotes $ \sinh(2 K)$, and where~\cite{or-ni-gu-pe-01b}
\begin{gather}
\hat{C}^{j}(M, N) =
 {1 \over {j!}}
\int_{-\pi}^{\pi} {{d\phi_1} \over {2 \pi}} \cdots
\int_{-\pi}^{\pi} {{d\phi_j} \over {2 \pi}}
\left( \prod_{n=1}^{j} {{1} \over {\sinh \gamma_n}} \right) \nonumber \\
\phantom{\hat{C}^{j}(M, N) =}{} \times
\left( \prod_{1 \le i \le k \le j} h_{ik} \right)^2
 \left( \prod_{n=1}^{j} x_n \right)^M
\cos\left( N \sum_{n=1}^{j} \phi_n \right)\label{involved}
\end{gather}
with
\begin{gather*}
x_n =
s + {{1} \over {s}} -\cos \phi_n
 -\left( \left( s + {{1} \over {s}} -\cos \phi_n \right)^2 -1 \right)^{1/2}, \\
 \sinh \gamma_n =
\left( \left( s + {{1} \over {s}} -\cos \phi_n \right)^2 -1 \right)^{1/2}, \qquad
 h_{ik} = {{ 2 (x_i x_k)^{1/2}
 \sin((\phi_i - \phi_k)/2) } \over {1 -x_i x_k }} .
\end{gather*}

For $ T < T_c$, let us f\/irst recall equation (3.15)
of Wu's paper~\cite{wu}, which reduces, for the
 diagonal correlations $ C(N,N)$,
to
\begin{gather}
(1-t)^{-1/4}  C(N,N)
 \sim
 1+ {{1} \over {(2 \pi)^2}}
 \int d\xi  \xi^N
\left((1-\alpha_2 \xi) (1-\alpha_2/\xi) \right)^{-1/2} \nonumber\\
\phantom{(1-t)^{-1/4}  C(N,N)  \sim}{} \times \int d{\xi'}  {\xi'}^{-N}
\left((1-\alpha_2 {\xi'}) (1-\alpha_2/{\xi'}) \right)^{1/2}
 {{ 1} \over {({\xi'} - \xi)^2 }}, \label{315}
\end{gather}
where $\alpha_2$ is $ t^{1/2}$.
Comparing with (\ref{formm}) we see that the second term in~(\ref{315}) is $ f^{(2)}_{N,N} = F^{(2)}_{N,N}$.

Performing the change of variables $ \xi = z_1$
and $ {\xi'} = 1/z_2$,
deforming the contour of integration for both $ z_1$ and $ z_2$
(one has to consider only the discontinuity
across the branch cut\footnote{For $ T<T_c$,
 $ \alpha_2 = t^{1/2} < 1$.}
running from $ 0$ to $ \alpha_2$),
and rescaling $ z_1$ and $ z_2$, in, respectively,
 $ x_1 = z_1/\alpha_2 $
and $ x_2 = z_2/\alpha_2$, we obtain:
\begin{gather*}
f^{(2)}_{N,N}(t) =
 F^{(2)}_{N,N}(t) =
 {t^{(N+1)}\over
 \pi^2} \int_0^1 x_1^N
dx_1 \int_0^1 x_2^N dx_2 \nonumber \\
\phantom{f^{(2)}_{N,N}(t) =
 F^{(2)}_{N,N}(t) =}{}  \times \left({x_{1}(1-x_{2})(1 -t x_{2})\over
 x_{2}(1-x_{1})(1 -t x_{1})}\right)^{1/2}
 (1 -t x_{1} x_{2})^{-2}.
\end{gather*}

Similarly, when $ T >T_c$, the leading term for $ G_{N, N}^{(1)} $
is given by equation (2.29) of~\cite{wu}
\begin{gather*}
 f_{N, N}^{(1)} = G_{N, N}^{(1)} =
{{-1} \over {2 \pi i}}
 \int_{C} dz {{ z^{N-1} } \over {
\left( (1 - t^{1/2} z )(1 - t^{1/2} z^{-1}) \right)^{1/2} }}
\end{gather*}
which, after deforming the contour of integration to the branch cut,
and scaling $ z = t^{1/2} x $, becomes
\begin{gather}
f_{N,N}^{(1)}(t) = G_{N,N}^{(1)}(t) = {{t^{N/2}} \over {\pi}}
 \int_0^{1} x^{N-1/2} (1-x)^{-1/2} (1 -x t)^{-1/2} dx
\nonumber\\
\phantom{f_{N,N}^{(1)}(t) = G_{N,N}^{(1)}(t)}{} =
 t^{N/2}  {(1/2)_N\over N!}
 {_2}F_1\left({1\over 2},N+{1\over 2}; N+1;t \right),\label{1}
\end{gather}
where $ {_2}F_1(a,b;c;z) $ is the
 hypergeometric function~\cite{Erde}.

When the low temperature
expansion of Section~3 of Wu~\cite{wu} is performed
to all orders, we f\/ind that (\ref{cm}) holds with
\begin{gather*}
F^{(2n)}_{N,N} =
{(-1)^{n+1} \over n}{1 \over (2\pi)^n}\int \prod_{j=1}^{2n}
{z_j^N dz_j \over 1 -z_jz_{j+1}}  \prod_{j=1}^{n}\left({
 (1-\alpha_2 z_{2j})(1-\alpha_2/z_{2j}) \over
(1 -\alpha_2 z_{2j-1})(1 -\alpha_2/z_{2j-1})}\right)^{1/2}
\end{gather*}
from which, after deformation of integration contours and rescaling,
one obtains, for $ T < T_c$,
the following {\em new integral representation}
of $ F^{(2n)}_{N,N}(t)$:
\begin{gather} F_{N,N}^{(2n)}(t) =
 {(-1)^{n+1} t^{n(N+1)}\over n \pi^{2n}} \! \int_{0}^1
 \prod_{j=1}^{2n}{ x_j^N dx_j \over 1 -t x_j x_{j+1}}
\prod_{j=1}^{n} \left({x_{2j-1}(1 -x_{2j})(1-tx_{2j})
\over x_{2j}(1 -x_{2j-1})(1 -t x_{2j-1})}\right)^{1/2}\!\! .\label{Fn}\!\!\!
\end{gather}

Similarly for $ T > T_c $ the expansion of Section 2 of Wu~\cite{wu}
is performed to
all orders and we f\/ind that (\ref{cp}) holds with $ F^{(2n)}_{N,N} $
 given by (\ref{Fn}) and
\begin{gather*}
G^{(2n+1)}_{N,N} =
 (-1)^n {1\over (2\pi)^{2n+1}}
 \int \prod_{j=1}^{n+1}(z_{j}^{N+1} dz_{j} )
 {{ 1} \over {z_1 z_{2n+1} }}
 \prod_{j=1}^{2n}{1\over 1 -z_jz_{j+1}}\nonumber\\
\phantom{G^{(2n+1)}_{N,N} =}{} \times \prod_{j=1}^{n+1}\left(
(1 -\alpha_2^{-1}z_{2j-1})
(1 -\alpha_2^{-1}/z_{2j-1})\right)^{-1/2}
 \prod_{j=1}^{n}
\left((1 -\alpha_2^{-1}z_{2j})
(1 -\alpha_2^{-1}/z_{2j})\right)^{1/2} .
\end{gather*}
Changing variables and deforming contours, we obtain:
\begin{gather*}
G^{(2n+1)}_{N,N}(t) = (-1)^{n} {t^{N(2n+1)/2+2n} \over
 \pi^{2n+1}}
 \int_{0}^{1} \prod_{j=1}^{2n+1}(x_{j}^{N+1} dx_{j})
 {{1} \over {x_1 x_{2n+1}}}
\prod_{j=1}^{2 n}{1\over 1 -t x_j x_{j+1}}\nonumber \\
\phantom{G^{(2n+1)}_{N,N}(t) =}{} \times \prod_{j=1}^{n+1}\left({x_{2j-1}\over
 (1 -x_{2j-1})(1 -t x_{2j-1})}\right)^{1/2}
 \prod_{j=1}^{n} \left(
 (1 -x_{2j})(1 -t x_{2j})/x_{2j}\right)^{1/2} . \nonumber
\end{gather*}

The form factor expressions are then obtained by expanding the exponentials.
Thus we f\/ind, for $ T<T_c$, that the form factors in (\ref{formm1}) read
\begin{gather}
f_{N,N}^{(2n)}(t) = {{ t^{n(N+n)}} \over {
 (n!)^2 }} {{1 } \over {\pi^{2n} }}
\int_0^1\prod_{k=1}^{2n} x_k^N dx_k  \prod_{j=1}^n
 \left({x_{2j-1}(1-x_{2j})(1 -t x_{2j})\over
x_{2j}(1-x_{2j-1})(1 -t x_{2j-1})}\right)^{1/2} \nonumber\\
\phantom{f_{N,N}^{(2n)}(t) =}{}  \times
 \prod_{j=1}^{n}
\prod_{k=1}^{n}(1 -t x_{2j-1} x_{2k})^{-2}
\prod_{1\leq j<k\leq n}(x_{2j-1} -x_{2k-1})^2(x_{2j} -x_{2k})^2,\label{2n}
\end{gather}
and, for $ T>T_c $, the odd form factors in (\ref{formp1}) read
\begin{gather}
f_{N,N}^{(2n+1)}(t) = t^{((2n+1)N/2 +n(n+1))}
 {{ 1 } \over {\pi^{2n+1} }}
 {{ 1 } \over {n! (n+1)! }} \nonumber\\
\phantom{f_{N,N}^{(2n+1)}(t) =}{}  \times \int_0^1\prod_{k=1}^{2n+1} x_k^{N} dx_k
\prod_{j=1}^{n+1} \left((1-x_{2j})
(1 -t x_{2j}) x_{2j}\right)^{1/2} \nonumber\\
\phantom{f_{N,N}^{(2n+1)}(t) =}{} \times
\prod_{j=1}^{n+1}
\left((1-x_{2j-1})(1 -t x_{2j-1}) x_{2j-1}\right)^{-1/2}
 \prod_{j=1}^{n+1} \prod_{k=1}^{n}
(1 -t x_{2j-1} x_{2k})^{-2} \nonumber \\
\phantom{f_{N,N}^{(2n+1)}(t) =}{} \times \prod_{1\leq j<k\leq n+1}(x_{2j-1}-x_{2k-1})^2
\prod_{1\leq j<k\leq n}(x_{2j}-x_{2k})^2,\label{2n-1}
\end{gather}
where the last product in (\ref{2n-1}) has to
be taken to be equal to unity for $ n= 0,1$.
We note that the factors $ 1/(n!)^2 $ and $ 1/(n! (n+1)!)$
in (\ref{2n}) and (\ref{2n-1}), arise because the integrands are {\em symmetric
functions of the variables} $ x_{2j}$ and $ x_{2j-1}$, {\em separately}. This
is to be contrasted with~(\ref{involved}), where there is no separation in the
odd and even integrals $ \phi_{j}$.
In the simplest case the previous integral representation (\ref{2n-1})
gives $ f_{N,N}^{(1)}(t)$ def\/ined by (\ref{1})
where one recognizes the Euler representation of an hypergeometric function.

Do note that the $ (G^{(2n+1)}_{N,N}, F^{(2n)}_{N+1,N+1})$ decomposition
in (\ref{cp}) {\em is not unique}. In contrast, the form factor expressions
(\ref{2n}), (\ref{2n-1}) are unique and well-def\/ined.

\section[Fuchsian linear differential equations for $ f_{N,N}^{(j)}(t)$]{Fuchsian linear dif\/ferential equations for $\boldsymbol{f_{N,N}^{(j)}(t)}$}
\label{holo}

We use formal computer algebra to
study the functions $ f^{(j)}_{N,N}$. We
 obtain the Fuchsian linear dif\/ferential
equations satisf\/ied by the $ f^{(j)}_{N,N}$
 for f\/ixed $ j \le 9$ and arbitrary $ N$.
We also f\/ind the truly remarkable result that the
 $ f^{(2j+1)}_{N,N}$ and $ f^{(2j)}_{N,N}$ are each
solutions of linear dif\/ferential operators which have
 a nested ``Russian-doll'' structure. Beyond this ``Russian doll''
structure, each linear dif\/ferential
operator is the {\em direct sum} of linear dif\/ferential operators
equivalent\footnote{For the notion of
equivalence of linear dif\/ferential
operators, see~\cite{Singer,PutSinger,hoeij2}.}
to symmetric powers of the
 second order dif\/ferential operator corresponding
 to $ f^{(1)}_{N,N}$, (or equivalently to the
 second order dif\/ferential operator $ L_E$,
 corresponding to the complete elliptic
integral $ E$).
A~direct consequence is that the form factors
$ f^{(2j+1)}_{N,N}$, and $ f^{(2j)}_{N,N}$
are {\em polynomials in the complete elliptic integrals of
the first and second kinds}, $ K$ and $ E$
\begin{gather}
\label{EK}
K = {_2}F_1 \left( 1/2, 1/2; 1; t \right), \qquad
E = {_2}F_1 \left( 1/2, -1/2; 1; t \right).
\end{gather}
A simple example is
$ f^{(2)}_{0,0} = K   (K-E)/2$.

In previous studies on the Ising
susceptibility~\cite{ze-bo-ha-ma-04,ze-bo-ha-ma-05b,ze-bo-ha-ma-05c,ze-bo-ha-ma-05},
ef\/f\/icient programs
were developed which, starting from long series expansions of a
holonomic function, produce the
linear ordinary dif\/ferential equation (in this case Fuchsian)
 satisf\/ied by the function. In order for these
programs to be used to study the $ f^{(j)}_{N,N}$'s we need to ef\/f\/iciently
produce long (up to several thousand terms) series expansions in $t$ of the
$ f^{(j)}_{N,N}$'s. We have done this by use of both the integral
representations (\ref{2n}), (\ref{2n-1}) and the representations
of $ f^{(j)}_{N,N}$ in terms of
theta functions of the {\em nome of elliptic functions},
presented in~\cite{or-ni-gu-pe-01b}.

We have obtained the Fuchsian linear dif\/ferential
 equations satisf\/ied by the (diagonal)
form factors $ f^{(j)}_{N,N}$ up to $ j = 9$.
The analysis of these linear dif\/ferential operators
 shows a remarkable Russian-doll structure
similar to the nesting of (the dif\/ferential operators of)
 the ${\tilde \chi}^{(j)}$'s
 found
 in~\cite{ze-bo-ha-ma-04,ze-bo-ha-ma-05b,ze-bo-ha-ma-05c,ze-bo-ha-ma-05}.
Specif\/ically we f\/ind that the expressions
$ f^{(1)}_{N,N}$, $ f^{(3)}_{N,N}$, $ f^{(5)}_{N,N}$,
 $ f^{(7)}_{N,N}$ are
actually solutions of the linear ODE for $ f^{(9)}_{N,N}$,
and that $ f^{(0)}_{N,N}$, $ f^{(2)}_{N,N}$,
$ f^{(4)}_{N,N}$, $ f^{(6)}_{N,N}$ are actually solutions
of the linear ODE for $ f^{(8)}_{N,N}$.
In addition, we f\/ind that all the linear dif\/ferential operators
for the $ f^{(j)}_{N,N}$'s have a direct sum decomposition in operators
equivalent to symmetric powers of the linear dif\/ferential operator corresponding to
$ f^{(1)}_{N,N}$. Consequently,
 all the $ f^{(j)}_{N,N}$'s can also be
 written as {\em polynomials in terms of
 the complete elliptic integrals} $ E$ and $ K$.
The remainder of
this section is devoted to the presentation of these results.

\subsection[Fuchsian linear differential equations for $ f^{(2n+1)}_{N,N}$]{Fuchsian linear dif\/ferential equations for $\boldsymbol{f^{(2n+1)}_{N,N}}$}
\label{Fuchsf2n-1}

The linear dif\/ferential operator $ F_9(N)$ which annihilates
$ f^{(9)}_{N,N}$ has the
following factorized form
\begin{gather*}
F_9(N) = L_{10}(N)  L_8(N)  L_6(N)
 L_4(N)  L_2(N),
\end{gather*}
where the linear dif\/ferential operators $ L_r(N)$ are of order $ r$.
The f\/irst two operators read
\begin{gather}
\label{l2nn}
 L_2(N) = Dt^2
 + {\frac {2 t-1}{ \left( t-1 \right) t}}  Dt
-{{1} \over {4 t}} +{{1} \over {4 (t-1) }}
 - {\frac {{N}^{2}}{ 4 {t}^{2}}},
\\
 L_4(N) = L_{4,0} - N^2  L_{4,2}
+{\frac {9}{16}} {\frac {{N}^{4}}{{t}^{4}}},\nonumber
\end{gather}
with $ Dt=d/dt $ and:
\begin{gather*}
 L_{4,0} = Dt^4
+10 {\frac { (2 t-1) }{ \left( t-1 \right) t}}  Dt^3
+ {\frac { \left( 241 {t}^{2}-241 t+46 \right)}
{ 2\quad \left( t-1 \right)^2 t^2 }}  Dt^2\nonumber \\
\phantom{L_{4,0} =}{} + {{ \left( 2 t -1\right)
 \left( 122 {t}^{2}-122 t+9 \right) }\over {
 \left( t-1 \right)^{3} t^3 }}  Dt
+{\frac {81}{16}} {\frac { \left( 5 t-1 \right)
 \left( 5 t-4 \right) }{{t}^{3} \left( t-1 \right)^{3}}} ,\nonumber
\\
 L_{4,2} = {\frac {5}{2}} {\frac { Dt^2}{ t^2}}
- {\frac { \left(23 - 32 t \right) }
{ 2 \left( t-1 \right) t^3 }}  Dt
-{\frac {9}{8}} {\frac {8-17 t}{ \left(t-1 \right) t^{4}}}.
\end{gather*}
The expressions (or forms) of $ L_6(N)$, $ L_{8}(N)$
and $ L_{10}(N)$ are given in \cite{Holo}. The
linear dif\/ferential operators $F_{2n+1}(N)$, which annihilate
$f^{(2n+1)}_{N,N}$ for $n=0, \dots, 3$, are such that:
\begin{gather}
F_7(N) =
 L_8(N)  L_6(N)  L_4(N)  L_2(N) , \qquad
F_5(N) = L_6(N)  L_4(N)  L_2(N) , \nonumber \\
F_3(N) = L_4(N)  L_2(N) , \qquad
F_1(N) = L_2(N) . \label{F7531}
\end{gather}
Thus we see that the linear dif\/ferential operator for $ f^{(2n-1)}_{N,N}$
{\em right divides} the linear dif\/ferential operator for $ f^{(2n+1)}_{N,N}$ for
 $ n \leq 3$. We conjecture
 that this property holds for all values of $n$.
We thus have a ``Russian-doll'' (telescopic) structure of
these successive linear dif\/ferential operators.

\subsection[Fuchsian linear differential equations for $ f^{(2n)}_{N,N}$]{Fuchsian linear dif\/ferential equations for $\boldsymbol{f^{(2n)}_{N,N}}$}
\label{Fuchsianf2n}

The linear dif\/ferential
 operator $ F_8(N)$ (corresponding to $ f^{(8)}_{N,N}$)
 has the following factorized form
\begin{gather*}
F_8(N) =  L_9(N) L_7(N)  L_5(N)  L_3(N)  L_1(N),
\end{gather*}
where the linear dif\/ferential operators $ L_r(N)$ are of order $ r$.
The f\/irst two read:
\begin{gather*}
 L_1(N) = Dt, \\
 L_3(N) = Dt^3
+4  {\frac { \left( 2 t-1 \right) }{ \left( t-1 \right) t }} Dt^2
+{\frac { \left(2 -15 t +14 {t}^{2} \right) }
{ \left( t-1 \right)^{2} t^{2}}} Dt
+ {\frac {8 {t}^{2}-15 t+5}
{ 2 \left( t-1 \right)^3 t^2 }}
-\left({\frac {{  Dt}}{{t}^{2}}}
 +{{1} \over {t^3}} \right) N^2 .
\end{gather*}
The expressions (or forms) of the linear dif\/ferential operators
$ L_5(N) $, $ L_7(N) $ and $ L_9(N) $
are given in \cite{Holo}.

Similarly to (\ref{F7531}) there is also a Russian-doll (telescopic) structure of
 these successive linear dif\/ferential operators:
\begin{gather}
 F_6(N) =
L_7(N)   L_5(N)   L_3(N)   L_1(N) , \qquad
 F_4(N) = L_5(N)   L_3(N)   L_1(N) , \nonumber \\
  F_2(N) = L_3(N)   L_1(N) , \qquad
  F_0(N) = L_1(N) .\label{F642}
\end{gather}
Again, we see that the linear dif\/ferential
operator for $ f^{(2n-2)}_{N,N}$
{\em right divides} the linear dif\/ferential
 operator for $ f^{(2n)}_{N,N}$ for
 $ n \leq 4$. We conjecture that
 this property holds for all values of $ n$.

\subsection{Direct sum structure}
\label{beyond}

Not only do the linear dif\/ferential operators $L_{j}(N)$ have a factorized
Russian-doll structure, but we have found that they also have a {\em direct
sum decomposition when the integer $N$ is fixed}.
To illustrate this direct sum decomposition, let us write
the corresponding linear dif\/ferential operator for $ f^{(3)}_{N,N}$
\begin{gather*}
F_3(N) = L_4(N)   L_2(N)
 = M_4(N) \oplus L_2(N),
\end{gather*}
where $ L_2(N)$ is the linear dif\/ferential operator for $ f^{(1)}_{N,N}$
and where the fourth order operator~$ M_4(N) $
is displayed in \cite{Holo}
for successive values of $ N$.
One remarks on these successive expressions
that the degree of each polynomial occurring in these linear
 dif\/ferential opera\-tors~$ M_4(N)$ {\em grows linearly with} $ N$.

As a further example consider $ f^{(5)}(N,N)$,
where we f\/ind that the corresponding linear
 dif\/ferential operator decomposes as
\begin{gather*}
F_5 = L_6(N)   L_4(N)   L_2(N)
 = M_6(N) \oplus M_4(N) \oplus L_2(N),
\end{gather*}
where $ L_2(N)$ is the linear dif\/ferential operator for $ f^{(1)}_{N,N}$,
$ M_4(N)$ is the previous fourth order dif\/ferential operator,
and the sixth order operator $ M_6(N) $
has again coef\/f\/icients whose degrees {\em grow
with $ N$} for successive values of $ N$.
There is nothing specif\/ic to $ f^{(3)}_{N,N}$ and $ f^{(5)}_{N,N}$:
similar results hold for all the $ f^{(n)}_{N,N}$'s,
$ n$ being even or odd.

In contrast with the Russian-doll way of writing the linear dif\/ferential
 operators for $ f^{(n)}_{N,N}$, the direct sum structure,
as a consequence of this growing degree, cannot, for generic $N$, be
written in a closed form as operators with polynomials coef\/f\/icients
in front of the derivatives.
This ``non-closure'' of the direct sum structure will have some
consequences when performing the {\em scaling limit} of
these linear dif\/ferential operators (see Section~\ref{scal} below).

\subsection[Equivalence of various $ L_j(N) $'s and
 $ M_j(N) $'s linear differential operators]{Equivalence of various $\boldsymbol{L_j(N)}$'s and
 $\boldsymbol{M_j(N)}$'s linear dif\/ferential operators}

\label{equiv2}

We f\/ind that the symmetric square\footnote{The symmetric $j$-th power of
a second order linear dif\/ferential operator
 having two solutions $ f_1$ and $ f_2$
is the linear dif\/ferential operator of order $j+1$, which has
 $ f_1^{j}, \dots , f_1^{j-k} f_2^{k} , \dots ,
 f_2^{j}$ as solutions. }
 of $ L_2(N)$
\begin{gather*}
{\rm Sym}^2(L_2(N)) =
 Dt^{3}
 +3 {\frac { \left(2 t -1\right) }{ \left( t-1 \right) t}}  Dt^2
 +{\frac { \left( 1-7 t+7 {t}^{2} \right) }
{ \left( t-1 \right)^{2}{t}^{2}}} Dt \nonumber \\
\phantom{{\rm Sym}^2(L_2(N)) =}{} -{{1} \over {2}} {\frac {1-2 t}
 { \left( t-1 \right)^{2}{t}^{2}}}
 -{{N^2} \over {t}}  Dt
 -{{N^2} \over {(t-1) t^2}}
\end{gather*}
and the linear dif\/ferential operator $ L_3(N) $
are equivalent\footnote{For the
equivalence of linear dif\/ferential
operators, see~\cite{Singer,PutSinger,hoeij2}.}
\begin{gather*}
L_3(N)   U(N) = V(N)   {\rm Sym}^2(L_2(N))
\end{gather*}
with the following intertwiners:
\begin{gather*}
U(N) =
 \left(t -1 \right)  t  Dt^{2}
+ \left( 3 t-1 \right)  Dt +1
+{\frac { \left( 1-t \right) }{t}} {N}^{2},
\\
 V(N) =
\left(t -1 \right)  t  Dt^{2}
+ \left( 11 t-5 \right)  Dt
 +{\frac { \left( 5 t-1 \right)
 \left( 5 t-4 \right) }{ \left( t-1 \right) t }}
 -{\frac { \left( t-1 \right)}{t}}  N^2 .
\end{gather*}

Similarly, with the symmetric cube of $ L_{2}(N)$,
we have the equivalence
\begin{gather*}
 L_4(N)  A(N) = B(N)   {\rm Sym}^3(L_2(N))
\end{gather*}
with:
\begin{gather*}
A(N) = ( t-1)  t  Dt^3
+ {{7 }\over {2}} \left( 2 t -1\right)  Dt^2
+ {\frac { \left( 41 {t}^{2}-41 t+6 \right)}
{ 4 \left( t-1 \right) t}}  Dt
 \nonumber \\
\phantom{A(N) =}{} + {\frac {9}{8}} {\frac {2 t-1}{ \left( t-1 \right) t}}
 -{\frac {9}{4}} {\frac { \left( t-1 \right)   N^2 }{t}}  Dt
-{\frac {9}{8}} {\frac { \left( 2 t -1\right) }{{t}^{2}}}  N^2 , \nonumber \\
B(N) = \left( t-1 \right) t   Dt^{3}
 +{{23} \over {2}} \left( 2 t-1 \right)   Dt^2
+{\frac {21}{4}} {\frac {
 \left( 6-29 t+29 {t}^{2} \right) }{ \left( t-1 \right) t}}  Dt\\
 \phantom{B(N) =}{}
 +{\frac {9}{8}} {\frac { \left( 2 t-1
 \right) \left( 125 {t}^{2}-125 t+16 \right) }
{ \left( t-1 \right)^{2} t^2}}
-{{9} \over {4}} {\frac { \left( t-1 \right) }{t}}
  N^2   Dt
-{\frac {9}{8}} {\frac { \left( 10 t-9 \right) }{{t}^{2}}}   N^2 .
\end{gather*}

More generally, all the $ L_m(N)$'s are
 $(m-1)$-sym\-met\-ric-power of $ L_2(N)$.
As a conse\-quen\-ce~their solutions are $(m-1)$-ho\-mo\-ge\-neous polynomials of the
two hypergeometric solutions of~$ L_2(N)$.

Similarly, for the linear dif\/ferential operators occurring in the direct sum,
one easily verif\/ies, for every integer $ N$,
that, for instance, the $ M_4(N)$'s are equivalent to the
 cubic-symmetric-power of~$ L_2(N)$
\begin{gather*}
 M_4(N)   Q(N) = S(N)   {\rm Sym}^3(L_2(N))
\end{gather*}
where, for $ N= 0, 1, 2$:
\begin{gather*}
Q(0) = \left(t-1 \right) t  Dt
 +t -{{1} \over {2}},\\
Q(1) = 2 \left( t-1 \right)^{3} t^2  Dt^{3}
+3 \left( 3-7 t+4 {t}^{2} \right)
 \left( t-1 \right) t  Dt^{2}
\nonumber \\
\phantom{Q(1) =}{} + \left( 12 {t}^{3}-28 {t}^{2} +{\frac {41}{2}} t
-{{9} \over {2}} \right)   Dt
+{{3} \over {4}}  {\frac{2 {t}^{2} -2 t+1}{t}},\nonumber\\
 Q(2) =
{{1} \over {3}} \left( t-1 \right)^{3}
 \left( 3+8 t+3 {t}^{2} \right) t   Dt^{3}  +{{1} \over {2}}
 \left( 15-t-35 {t}^{2}+15 {t}^{3}+6 {t}^{4} \right)
\left( t -1 \right)  Dt^{2} \nonumber \\
\phantom{Q(2) =}{} -{{1} \over {24}} {\frac { \left( 18 {t}^{5}
-12 {t}^{4}-97 {t}^{3}+577 {t}^{2}-738 t+252
 \right) }{t} }   Dt \nonumber \\
\phantom{Q(2) =}{}
-{{1} \over {16}} {\frac {12 {t}^{5}+14 {t}^{4}
-260 {t}^{3}+497 {t}^{2}-314 t+24}{{t}^{2}}} .
\end{gather*}

As a further example, one can verify,
for every value of the integer $ N$,
that the sixth order operator $ M_6(N) $ is equivalent
to the f\/ifth symmetric power of $ L_2(N)$.
The solutions of the linear dif\/ferential operators
$ M_m(N)$ are also $(m-1)$-homogeneous polynomials of the
two hypergeometric solutions of~$ L_2(N)$.
As a consequence of this direct sum decomposition,
 the solutions~$ f^{(n)}(N, N)$
are (non-homogeneous) {\em polynomials of the
two hypergeometric solutions of}~$ L_2(N)$ or,
 equivalently, $ f^{(1)}_{N,N}$ (or the hypergeometric
solution of (\ref{l2nn})) and its
f\/irst derivative. The second order linear
 dif\/ferential operator $ L_2(N)$
is equivalent~\cite{PainleveFuchs} to the
second order linear dif\/ferential operator $ L_E$
\begin{gather*}
L _E = 4 t   Dt^{2}
 + 4 Dt - {{1} \over {t-1}}
\end{gather*}
 corresponding to the complete elliptic integral of the second kind $ E$. As a consequence
of the previously described direct sum decomposition, the $ f^{(n)}_{N,N}$'s
can also be written as {\em polynomial expressions of the
 complete elliptic integral} of the second kind $ E$ and its f\/irst derivative
 $ E'$, or alternatively, $ E$ and
 {\em the complete elliptic integral\/}\footnote{Of course no confusion
is possible between the
complete elliptic integral $ K$ and the usual Ising model temperature variable sometimes
denoted $K= J/kT$.} of the f\/irst kind $ K$.

Let us just give here a set of miscellaneous
examples of polynomial expressions
of various form factors. For $ f^{(2)}_{N,N}$, one has
\begin{gather*}
2 f^{(2)}_{0,0} =
 \left(K -E \right)  K , \qquad
 2 f^{(2)}_{1,1} =
1 -3 K E - \left(t-2 \right)  {K}^{2} , \nonumber \\
 6 t  f^{(2)}_{2,2} =
6 t - \left( 2+6 {t}^{2}-11 t \right)   {K}^{2}
 - \left( 15 t -4 \right)  K E
 -2 \left( 1+t \right)  {E}^{2}, \nonumber \\
 90 t^2   f^{(2)}_{3,3} = 135 {t}^{2}
- \left( 137 {t}^{3}-242 {t}^{2}+52 t+8 \right)  {K}^{2}  \nonumber \\
 \phantom{90 t^2   f^{(2)}_{3,3} =}{}  + \left( 8
{t}^{3}-319 {t}^{2}+112 t+16 \right)  K E -4 \left( 1+t \right)
 \left( 2 {t}^{2}+13 t+2 \right) {E}^{2}, \nonumber \\
 3150 t^3  f^{(2)}_{4,4}
= 6300 {t}^{3}  - \left( 32 {t}^{5}+2552 {t}^{2}
 +128+6440 {t}^{4}-11191 {t}^{3}
+464 t \right)  {K}^{2} \nonumber \\
\phantom{ 3150 t^3  f^{(2)}_{4,4}=}{}  + \left( 128 {t}^{5}+5648 {t}^{2}-14519 {t}^{3}
+1056 t +576 {t}^{4}+256 \right)  E K \nonumber \\
\phantom{ 3150 t^3  f^{(2)}_{4,4}=}{}  -8 \left( 1+t \right)
 \left(16 {t}^{4}+58 {t}^{3}+333 {t}^{2}
 +58 t+16 \right)  {E}^{2}, 
\end{gather*}
where $ E$ and $ K$ are given by (\ref{EK}).
Other examples are given in~\cite{Holo}.

\medskip

\noindent
{\bf Miscellaneous remarks.} All these remarkable structures {\em are not
 restricted to diagonal two-point correlation functions}. We keep on restricting
to the {\em isotropic} Ising model: for the ani\-sotro\-pic Ising model one has (for the correlations
 and may have for the form factors) similar
but more complicated results involving the complete
elliptic integral of the third kind $\Pi$ (see for instance equation~(3.35) in
H.~Au-Yang and J.H.H.~Perk~\cite{Perk4}, or pp.~23--48
in~\cite{Perk5}, more recently~\cite{Witte} and for a sketch
of how the algebro-dif\/ferential structures
 generalize in that anisotropic case~\cite{Fuchs}).

$ \bullet$ Further, one can calculate various
$ j$-particle contributions $ f^{(j)}_{M,N}$
of the {\em off-diagonal} two point correlation functions,
and verify, again, that they are, in the isotropic case, {\em also polynomial
expressions of the complete elliptic integrals}
$ E$ and $ K$. For instance:
\begin{gather*}
C^{(2)}(0, 1) =
{{3} \over {8}}
 - {{1} \over {4}} \left( 1+{s}^{2} \right) K
-{{1} \over {2}} E K
 -{{1} \over {8}} \left( {s}^{2}-3 \right)
 \left(1+{s}^{2} \right)  K^{2}, \nonumber
\end{gather*}
where $ s = \sinh(2 K)$.
Other miscellaneous examples of such of\/f-diagonal
 $ j$-particle contributions
are displayed in~\cite{Holo}. In the {\em anisotropic case} polynomial expressions
of $ E$ and $ K$ and complete
elliptic integral of the third kind $\Pi$ could
 take place for $ j$-particle contributions $ f^{(j)}_{M,N}$.
The occurrence of elliptic integral of the third kind and not more involved
hyperelliptic integrals is still not clear
(see after equation~(3.20) in~\cite{or-ni-gu-pe-01b} the remark
 on Glasser's unpublished work). This work is still in progress.

$ \bullet$ The products of the two-point
 correlation functions $ C(N, N)$
are also solutions of Fuchsian linear ODE's. As a consequence the
equal-time xx-correlations~\cite{Capel} of the
 free-fermion zero-f\/ield $ XY$ {\em quantum chain}, which are,
 alternatingly, $ C(N,N)^2$ and
$ C(N, N) C(N+1, N+1)$, also satisfy a Fuchsian linear ODE.

$ \bullet$ Far beyond, recalling Boel, Kasteleyn and Groeneveld
 papers~\cite{Groeneveld2,Groeneveld3,Groeneveld}
one can see that all the two-point correlation functions
of Ising models ({\em not necessarily free-fermion} Ising models!)
can be expressed as sums, weighted with $ \pm$ signs,
of products of two-point correlation functions. Consequently
{\em all the $ n$-point correlation functions of the square Ising model
are (simple) polynomial expressions of the
complete elliptic integrals $ E$ and $ K$}
and, of course, the $ n$-point correlation
 functions of the square Ising model
are solutions of Fuchsian linear ODE's. For the anisotropic Ising model
the $ n$-point correlation
 functions are solutions of PDE's associated with complete elliptic integrals
of the third kind (see~\cite{Fuchs} for a sketch).

$ \bullet$ Recalling the relations
(\ref{formm}), (\ref{formp}) between the
two-point correlation functions and the form
factors we see that, since the isotropic two-point correlation functions and the form
factors are both polynomial expressions of the
 complete elliptic integral $ E$ and $ K$,
 relations (\ref{formm}), (\ref{formp}) can be interpreted as an inf\/inite
number of quite {\em non trivial identities} on the
 complete elliptic integral $ E$ and $ K$,
 for instance:
\begin{gather*}
C_{+}(N,N) = P_1^{(N, N)}(E, K) =
(1-t)^{1/4}  \sum_{n=0}^{\infty} f^{(2n+1)}_{N,N}
 = (1-t)^{1/4}
 \sum_{n=0}^{\infty} Q^{(2n+1)}_{N,N}(E, K).
\end{gather*}
We have similar identities for the (isotropic) of\/f-diagonal two-point correlations
$ C(M, N)$. These linear relations on an inf\/inite number of
polynomial expressions of the
 complete elliptic integrals\footnote{In the anisotropic case we could have
identities on an inf\/inite number of
polynomial expressions of the
 complete elliptic integrals of the f\/irst, second
 and third kind (and hyperelliptic integrals \dots?). This work is still in progress.}~$ E$ and~$ K$ have to be compared
with the inf\/inite number of (non-linear) relations on a~f\/inite number of
polynomial identities on the
 complete elliptic integral $ E$ and $ K$ which
correspond to~(\ref{perk}), the quadratic f\/inite
dif\/ference relations~\cite{Perk3,mcc-wu-80,Perk2,PerkDunk} on the
two-point correlation functions displayed in Appendix~\ref{painl}.

$ \bullet$ At criticality, $ k = 1$, many remarkable
and much simpler identities can be obtained,
for instance the formula\footnote{This formula has f\/irst been obtained
in~\cite{Perk6} at the critical temperature $ k = 1$.} (2.34) in~\cite{Perk} on
the next to the diagonal (anisotropic) two-point
correlations (see also~\cite{Witte}):
\begin{gather*}
C(N-1, N) = C(N, N)   \cosh(2 K)
 F\big(1/2, N; N +1/2, -\sinh(2K)^2\big),
\end{gather*}
where $ F$ is the hypergeometric function.

\subsection{The elliptic representation of Painlev\'e VI}
\label{elliptic}

The results we have underlined in this section, namely the unexpectedly
simple and remarkable polynomial expressions
for the form factors $ f^{(j)}_{N,N}$,
 correspond to the fact that the associated linear dif\/ferential operators
are direct sums of operators equivalent to
 symmetric powers of the second order dif\/ferential operator $ L_E$.
We already encountered this central key role played
 by the linear dif\/ferential operator $ L_E$,
or the hypergeometric second order linear dif\/ferential
operator~(46) given in \cite{PainleveFuchs},
in our previous holonomic analysis of the two-point correlation
functions of the Ising model~\cite{PainleveFuchs}.
In order to understand the key role played by $ L_E$,
or equivalently operator~$L_2(N)$, it is worth
 recalling (see~\cite{man}, or for a review~\cite{Guzzetti}) the so-called
``{\em elliptic representation}'' of
 Painlev\'e~VI. This elliptic representation
of Painlev\'e~VI amounts to seeing Painlev\'e~VI as a~``deformation'' (see equation~(33) in~\cite{Guzzetti}) of the
hypergeometric linear dif\/ferential equation
 associated with the linear dif\/ferential operator
\begin{gather*}
{\cal L} = (1-t) t   Dt^2 + (1-2 t)   Dt -{{1} \over {4}}.
\end{gather*}
One easily verif\/ies that this linear dif\/ferential operator is actually
equivalent (in the sense of the equivalence
of dif\/ferential operators) with $ L_E$, or equivalently $L_2(N)$.
This deep relation between {\em elliptic curves and Painlev\'e~VI}
explains the occurrence of Painlev\'e~VI on the Ising model,
and on other lattice Yang--Baxter integrable models
which are canonically parametrized in term of elliptic functions
(like the eight-vertex Baxter model, the RSOS models,
see for instance~\cite{Manga}).
One can see in Section~6 of~\cite{Holo}, other examples of this deep connection
between the transcendent solutions
of Painlev\'e VI and the {\em theory of elliptic functions, modular curves
 and quasi-modular functions}.

Along this line one should note that other
linear dif\/ferential operators, not straightforwardly linked to $ L_E$
but more generally to the theory of elliptic functions
and {\em modular forms} (quasi-modular forms \dots),
 also emerge in the analysis of the
$ \lambda$-extensions of the two-point correlation
 functions of the Ising model, for
selected\footnote{For generic values
 of $ \lambda$, the $ \lambda$-extension
$ C(M, N; \lambda$) are not holonomic.} values of $ \lambda$:
 $\lambda=\cos(\pi m/n)$. This is detailed in Appendix~\ref{aa}.

\section[The scaling limit of $f^{(j)}_{N,N}$]{The scaling limit of $\boldsymbol{f^{(j)}_{N,N}}$}
\label{scal}

The closed (exact) formulae (\ref{F7531}), (\ref{F642})
we obtain for the
 linear dif\/ferential operators in these
nested ``Russian doll'' structures, enable us to take the
{\em scaling limit} of these linear operators.
We study this scaling limit in this section
and show that the ``Russian-doll'' structure remains valid.
The linear dif\/ferential operators
in that ``scaled'' nested Russian-doll structure remain equivalent to
 the symmetric power of a singled-out second order linear
dif\/ferential operator (corresponding to the modif\/ied Bessel function).
In contrast, in the scaling
 limit, {\em the direct sum of operators decomposition
structure is lost},
and we explain why.

The scaling of the $ f^{(n)}_{N,N}$'s amounts, on the functions,
and on the corresponding dif\/ferential operators, to
 taking the limit $ N \rightarrow \infty$
and $ t \rightarrow 1$, keeping the limit
$ x = N  (1-t)$ f\/inite, or in other words, to
 performing the change of variables $ t=1-x/N$,
keeping only the leading term in $ N$.
Performing these straightforward calculations, the linear
dif\/ferential operators in $ t$
for the $ f^{(n)}_{N,N}$'s where $ N$ was a parameter,
 become linear dif\/ferential operators in the only
scaling variable $ x$.

Calling $ F^{\rm scal}_j$ the scaling limit of the operator $ F_{j}(N)$ we
f\/ind for $j$ even that
\begin{gather}
F_{8}^{\rm scal} =
 L_9^{\rm scal}   L_7^{\rm scal}   L_5^{\rm scal}
  L_3^{\rm scal}   L_1^{\rm scal} , \qquad
F_{6}^{\rm scal} = L_7^{\rm scal}   L_5^{\rm scal}
  L_3^{\rm scal}   L_1^{\rm scal} , \nonumber \\
F_{4}^{\rm scal} = L_5^{\rm scal}
 L_3^{\rm scal}  L_1^{\rm scal} , \qquad
F_{2}^{\rm scal} = L_3^{\rm scal}  L_1^{\rm scal} , \qquad
F_{0}^{\rm scal} = L_1^{\rm scal} ,\label{f6scal}
\end{gather}
where ($Dx = d/dx$)
\begin{gather*}
 L_5^{\rm scal} = 2 {x}^{5} {{ Dx}}^{5}
+10 {x}^{4}{{ Dx}}^{4}-2 {x}^{3} \left( 7+5 {x}^{2} \right)
{{Dx}}^{3} \nonumber \\
\phantom{L_5^{\rm scal} =}{} +2 \left( -16+13 {x}^{2} \right) {x}^{2} {{ Dx}}^{2}
 +2 \left( 5-12 {x}^{2}+4 {x}^{4} \right) x { Dx}  -10 +8 {x}^{2} -24 {x}^{4},\nonumber \\
L_3^{\rm scal} = 2 {x}^{3} {{  Dx}}^{3} +8 {x}^{2}{{ Dx}}^{2}
-2 \left( x-1 \right) \left( x+1 \right) x { Dx}
-2, \nonumber \\
L_1^{\rm scal} = Dx
\end{gather*}
and $ L_9^{\rm scal}$, $ L_7^{\rm scal}$ are given in \cite{Holo}.

Similarly, for $j$ odd, we have
\begin{gather}
F_{9}^{\rm scal} =
L_{10}^{\rm scal}  L_8^{\rm scal}  L_6^{\rm scal}
L_4^{\rm scal}  L_2^{\rm scal} , \qquad
F_{7}^{\rm scal} = L_8^{\rm scal}  L_6^{\rm scal}  L_4^{\rm scal}  L_2^{\rm scal} ,\nonumber \\
F_{5}^{\rm scal} = L_6^{\rm scal}  L_4^{\rm scal} L_2^{\rm scal} , \qquad
F_{3}^{\rm scal} = L_4^{\rm scal}  L_2^{\rm scal} , \qquad
F_{1}^{\rm scal} = L_2^{\rm scal} ,\label{f7scal}
\end{gather}
where{\samepage
\begin{gather*}
L_4^{\rm scal} = 16 x^4   Dx^4 +96 x^3 Dx^3
+ 40 \left( 2-x^2 \right) x^2 Dx^2  +8 \left( {x}^{2}-2 \right) x Dx
+9 x^{4} -8 {x}^{2}+16,\nonumber \\
 L_2^{\rm scal} = 4 x   Dx^{2} +4 Dx -x
\end{gather*}
and $ L_{10}^{\rm scal} $, $ L_8^{\rm scal} $, $ L_6^{\rm scal} $
are given in \cite{Holo}.}

Thus, we see that the scaled operators $ F_j^{\rm scal}$ have a ``Russian-doll''
structure straightforwardly inherited from the one for the lattice operators $F_j(N)$.

Consider the linear second-order dif\/ferential
 operator corresponding to the modif\/ied
Bessel function $K_n(x/2)$ for $ n=0$, namely:
\begin{gather}
\label{modBessel}
 B = {{Dx}}^{2} +{\frac {{Dx}}{x}} - {{1} \over {4}}.
\end{gather}
We recognize, in this linear dif\/ferential
 operator, the exact identif\/ication with
the scaled dif\/fe\-ren\-tial operator $F_1^{\rm scal}=L_2^{\rm scal}$.
We f\/ind that the symmetric square of the linear
 dif\/ferential operator $ B$, and the scaled
operator $ L_3^{\rm scal} $ {\em are equivalent}:
\begin{gather*}
 L_3^{\rm scal}   \big( x {{Dx}}^{2} +2 {Dx} -x\big) =
 \big(2 {x}^{4}{{Dx}}^{2}+12 {x}^{3}{Dx}
-2 {x}^{4}+8 {x}^{2}\big) {\rm Sym}^2(B). \nonumber
\end{gather*}
Similarly, the symmetric third power of the
 linear dif\/ferential operator $ B$,
 and the scaled operator~$ L_4^{\rm scal} $ are equivalent,
and, more generally, the symmetric $j$-th power of
 (\ref{modBessel}) and the scaled operator~$ L_{j+1}^{\rm scal} $ {\em are equivalent}:
\begin{gather*}
L_{j+1}^{\rm scal} \simeq {\rm Sym}^j(B).
\end{gather*}

Recall that the linear dif\/ferential operators $F_j(N)$, corresponding
 to the form factors $f^{(j)}_{N,N}$, can
be written as direct sums {\em only when} the integer $ N$ is f\/ixed.
At the scaling limit, this feature disappears
for the scaled linear dif\/ferential operators
$ F_j^{\rm scal}$ which have no direct sums.
Therefore while the scaling limit
 preserves the Russian-doll (telescopic)
structure (see~(\ref{F7531}),~(\ref{f7scal})) and also preserves
 the fact that the various operators in this Russian-doll (telescopic)
structure are equivalent to symmetric
powers of an operator (\ref{modBessel})
which replaces the operator $ L_E$,
the {\em direct sum structure is lost}.
As a consequence the scaling of the $ f^{(j)}_{N,N}$'s
{\em cannot be seen as simple polynomials of modified Bessel functions}.

There is one exception that concerns $ f^{(2)}_{N,N}$. Its scaled linear
dif\/ferential operator $ F_{2}^{\rm scal} $, has the non shared property of
being equivalent to the {\em direct sum} of $ Dx$
 with the symmetric square of (\ref{modBessel}), namely:
\begin{gather*}
F_{2}^{\rm scal} =
 L_1^{\rm scal} \oplus L_3^{\rm scal} \simeq Dx \oplus
{\rm Sym}^2(B).
\end{gather*}
From this equivalence, one immediately deduces the expression of the scaling
of the $ f^{(2)}_{N,N}$ as a {\em quadratic expression of the modified
Bessel functions} of $ x/2$ which actually identif\/ies
with formula (2.31b)--(3.151) in~\cite{wu-mc-tr-ba-76}.

 The occurrence of modif\/ied Bessel functions, emerging from {\em a
confluence of two regular singularities} of the
 complete elliptic integrals $ E$ and $ K$, or from
the hypergeometric function $ _{2}F_1$, should not be considered
as a surprise if one recalls the following limit of
the hypergeometric function $ _{2}F_1$ yielding
conf\/luent hypergeometric functions
$ _{1}F_1$. These conf\/luent hypergeometric functions,
$ _{1}F_1$, are nothing but {\em modified Bessel functions}~\cite{Erde}
\begin{gather*}
 _{2}F_1\left(a, p, b; {{z} \over {p}}\right) \quad
\rightarrow \quad _{1}F_1(a, b; z)
\qquad \hbox{when} \quad
p \rightarrow  \infty, \nonumber \\
 I(\nu, z) =
 {{ z^{\nu} } \over {2^{\nu} e^z \Gamma(\nu+1) }}
 _{1}F_1\left(\nu +{{1} \over {2}}, 2 \nu +1; 2 z\right).
\end{gather*}

\noindent
{\bf Remark.} It was shown, in Section~\ref{holo}, as a consequence of
 the decomposition of their linear dif\/ferential
operators in direct sums of operators
equivalent to symmetric
 powers of $ L_E$, that the functions $f^{(n)}_{N,N}$ are
polynomial expressions of $ E$ and $ K$
functions. Therefore their singularities
are only the three {\em regular} points
$ t =0$, $ t =1$ and $ t =\infty$.
The scaling limit
 ($ t= 1-x/N$, $t \rightarrow 1$, $N \rightarrow \infty$)
 corresponds to the {\em confluence}
 of the two regular singularities $ t =0$ and $ t = \infty$, yielding
the, now, {\em irregular} singularity at $ x=\infty$.
The occurrence of irregular singularities with their Stokes phenomenon,
 and, especially, the {\em loss of a remarkable direct sum structure},
shows that the scaling limit is a quite non-trivial limit.

Contrary to the common wisdom, the scaling limit does not correspond
to more ``fundamental'' symmetries and structures (more universal \dots):
this limit {\em actually destroys most of the remarkable
 structures and symmetries of the
 lattice models}\footnote{This kind of results should not be
a surprise for the people working on integrable lattice models,
discrete dynamical systems,
or on Painlev\'e equations~\cite{Sakai1,Sakai2}.}.

\section{Bridging with other formula of form factors\\ in the scaling
 limit: work in progress}
\label{morebrid}

The Ising Form Factors in the scaling limit as they can be
found in Wu, McCoy, Tracy, Barouch~\cite{wu-mc-tr-ba-76} read
\begin{gather}
\label{wmtbform}
 f^{(2 n)} = (-1)^n {{1} \over { \pi^{2 n} n}}
\int_{1}^{\infty} dy_1 \cdots \int_{1}^{\infty} dy_{2n}   \prod_{j=1}^{2 n} {{e^{-t y_j} }
 \over { (y_j^2 -1)^{1/2} (y_j +y_{j+1})
}} \prod_{j=1}^{ n} (y^2_{2j} -1),
\end{gather}
and:
\begin{gather*}
 g^{(2 k+1)} = (-1)^k {{1} \over { \pi^{2 k+1}}}
\int_{1}^{\infty} dy_1 \cdots \int_{1}^{\infty} dy_{2 k +1}   \prod_{j=1}^{2 k+1}
 {{e^{-t y_j} } \over { (y_j^2 -1)^{1/2} }}
 \prod_{j=1}^{2 k} {{1} \over{ (y_j +y_{j+1})}}
 \prod_{j=1}^{k} (y^2_{2j} -1).
\end{gather*}
The Ising Form Factors in the
 scaling limit are also given, in
many f\/ield theory papers, as follows ($ y_j =
\cosh\theta_j$, see (9) in~\cite{Babelon}, see also Mussardo~\cite{Muss}):
\begin{gather}
\label{babel}
 C^{(n)}_{\pm} = {{(\pm)^n} \over {n!}}
\int_{-\infty}^{\infty} \cdots \int_{-\infty}^{\infty}
 \prod_{j=1}^{n}
\left( {{ d\theta_j} \over {2 \pi}}
 e^{-r \cosh\theta_j} \right)
\prod_{i < j}
 \tanh^2\left( {{ \theta_i -\theta_j} \over {2}} \right).
\end{gather}

It remains to show that those expressions are actually solutions of the
scaled linear dif\/ferential operators displayed in the previous Section~\ref{scal},
namely (\ref{f6scal}) and (\ref{f7scal}).
A straight check yields too large formal calculations. Our
strategy should rather be to obtain the series expansions
of the $n$-fold integrals
(\ref{wmtbform}) or (\ref{babel}),
in the $t$ variable for (\ref{wmtbform}), or
the $ r$ variable for (\ref{babel}), and check that
these series expansions are actually solutions of the
non-Fuchsian linear dif\/ferential operators
(\ref{f6scal}) and (\ref{f7scal}) of Section~\ref{scal}.

These checks will show that the expressions
(\ref{wmtbform}), (\ref{babel}) of the $n$-fold integrals
of the scaled form factors are actually solutions
of linear dif\/ferential operators with an
 {\em irregular singularity at infinity}
and with a remarkable Russian-doll structure
but no direct sum structure. This will
indicate that such expressions
(\ref{wmtbform}), (\ref{babel}) generalise
modif\/ied Bessel functions but {\em cannot be simply expressed
in terms of polynomial expressions of modified Bessel functions}.
The interpretation of these expressions in terms of $ \tau$-functions
(Hirota equations, hierarchies, \dots) and the link between these
Russian-doll structures and B\"acklund transformations or Hirota
transformations remains to be done in detail.

\section[Other $ n$-fold integrals: from diagonal
correlation functions to the
susceptibility of the Ising model]{Other $\boldsymbol{n}$-fold integrals: from diagonal
correlation functions\\ to the
susceptibility of the Ising model}
\label{fromn-fold}

The study of two-point correlation functions (even $ n$-points correlations \dots)
 can be seen as
a~``warm-up'' for the truly challenging problem
of the study of the {\em susceptibility of the Ising model}
and its associated $ n$-fold integrals,
 the $ \chi^{(n)}$ (see next Section~\ref{n-fold} below). Staying
close to the diagonal correlation functions
we have introduced a simplif\/ication of the susceptibility of the Ising model
 by considering a magnetic f\/ield restricted to one diagonal of the
square lattice~\cite{Diag}. For this ``diagonal susceptibility''
model~\cite{Diag}, we benef\/ited from
the {\em form factor decomposition} of the diagonal two-point correlations
$ C(N,N)$, that has been recently
presented~\cite{Holo}, and subsequently proved by
 Lyberg and McCoy~\cite{lyb-mcc-07}.
The corresponding $ n$-fold integrals $ \chi^{(n)}_d$
were found to exhibit remarkable direct sum structures inherited
from the direct sum structures of the form factor~\cite{Holo,Diag}.
The linear dif\/ferential operators of the form factor~\cite{Holo}
being closely linked to the second order
dif\/ferential operator $ L_E$ (resp.~$ L_K$)
of the complete elliptic integrals~$ E$ (resp.~$ K$),
this ``diagonal susceptibility''~\cite{Diag} is also closely linked to the
elliptic curves of the two-dimensional Ising model.
By way of contrast, we note that the singularities of the linear ODE's for
these \mbox{$n$-fold} integrals~\cite{Diag} $ \chi^{(n)}_d$
are quite elementary (consisting of only $n$-th roots of unity) in comparison
with the singularities we will encounter below with the quite simple integrals
 (\ref{chinaked}).

 Using the form factor expansions (\ref{formm1}) and
(\ref{formp1}), the $ \lambda$-extension
of this diagonal susceptibility can be written as
\begin{gather*}
k_B T  \chi_{d\pm}(t;\lambda) =
(1-t)^{1/4}   \sum_{j}^{\infty} \lambda^j
  {\tilde \chi}^{(j)}_{d\pm},
 \end{gather*}
where the sum is over $j$ even (resp.~odd) for $ T$
 below (resp.~above) $T_c$ and where
\begin{gather*}
{\tilde \chi}^{(j)}_{d\pm} =
\sum_{N=-\infty}^{\infty} f^{(j)}_{N,N}(t).
\end{gather*}
By use of the explicit expressions (\ref{2n}), and (\ref{2n-1}),
for $ f^{(j)}_{N,N}$ we f\/ind explicitly, for $ T<T_c$, that
\begin{gather*}
{\tilde \chi}^{(2n)}_{d-} = {{ t^{n^2}} \over {
 (n!)^2 }} {{1 } \over {\pi^{2n} }}
\int_0^1 \cdots \int_0^1\prod_{k=1}^{2n} dx_k
  {1 +t^n x_1\cdots x_{2n}\over
 1 -t^n x_1 \cdots x_{2n}}\prod_{j=1}^n
\left({x_{2j-1}(1-x_{2j})(1-tx_{2j})\over
x_{2j}(1-x_{2j-1})(1 -t x_{2j-1})}\right)^{1/2}
\nonumber \\
\phantom{{\tilde \chi}^{(2n)}_{d-} =}{}  \times
 \prod_{1 \leq j \leq n}
\prod_{1 \leq k \leq n}(1 -t x_{2j-1} x_{2k})^{-2}
\prod_{1 \leq j<k\leq n}(x_{2j-1}-x_{2k-1})^2 (x_{2j}-x_{2k})^2
\end{gather*}
and, for $ T>T_c$, that:
\begin{gather*}
{\tilde \chi}^{(2n+1)}_{d+} =
{{ t^{n(n+1))} } \over {n! (n+1)! }}  {{ 1 } \over {\pi^{2n+1} }}
\int_0^1 \cdots \int_0^1
\prod_{k=1}^{2n+1} x_k^{N} dx_k \nonumber\\
\phantom{{\tilde \chi}^{(2n+1)}_{d+} =}{}  \times {1 +t^{n+1/2} x_1\cdots x_{2n+1}\over
 1 -t^{n+1/2} x_1\cdots x_{2n+1}}
 \prod_{j=1}^{n+1} \left((1-x_{2j})(1 -t x_{2j})
 x_{2j}\right)^{1/2} \nonumber \\
\phantom{{\tilde \chi}^{(2n+1)}_{d+} =}{} \times\prod_{j=1}^{n+1}
\left((1 -x_{2j-1})(1 -t x_{2j-1})   x_{2j-1}\right)^{-1/2}
  \prod_{1\leq j\leq n+1}\prod_{1\leq k \leq n}
(1 -t x_{2j-1} x_{2k})^{-2} \nonumber \\
\phantom{{\tilde \chi}^{(2n+1)}_{d+} =}{}  \times
\prod_{1 \leq j<k\leq n+1}(x_{2j-1} -x_{2k-1})^2
\prod_{1\leq j<k\leq n}(x_{2j}-x_{2k})^2. 
\end{gather*}

We have also found~\cite{Diag}, for
$ j = 1, \dots , 4$, that the ${\tilde \chi}_{d\pm}^{(j)}$'s
satisfy Fuchsian linear dif\/ferential equations
which have a Russian-doll nesting just as
was found for the $ {\tilde \chi}_n$'s
 in~\cite{ze-bo-ha-ma-04,ze-bo-ha-ma-05b,ze-bo-ha-ma-05c,ze-bo-ha-ma-05}.
In the case of these $ j$-particle components of
 the ``diagonal'' susceptibility, we can see
that this Russian-doll nesting of the corresponding
linear dif\/ferential operators
is {\em straightforwardly inheri\-ted}~\cite{Holo}, not
 from the Russian-doll nesting
of the diagonal form factors $ f^{(j)}(N, N)$'s (this is not suf\/f\/icient), but
from their direct sum (of operators equivalent to symmetric powers)
 decomposition.

Direct sum decompositions of the $ {\tilde \chi}^{(2n)}_{d-} $
 and $ {\tilde \chi}^{(2n+1)}_{d+}$
are straightforwardly
inherited~\cite{Diag} from direct sums of the $f^{(j)}_{N,N}$ thus yielding
a scenario for the direct sum decompositions of the ``true''~$\chi^{(n)}$.
However, recalling the non-unicity of the
$ (G^{(2n+1)}_{N,N}, F^{(2n)}_{N,N})$
(see Section~\ref{new}), the direct sum decomposition of the~$\chi^{(2n)}$
can be seen as a canonical one, when the direct sum decomposition of
the $\chi^{(2n+1)}$ is not.

\section[Other $ n$-fold integrals linked to the
susceptibility of the Ising model]{Other $\boldsymbol{n}$-fold integrals linked to the
susceptibility\\ of the Ising model}
\label{n-fold}

The susceptibility $ \chi$ of the square lattice Ising model
has been shown by Wu, McCoy, Tracy and
 Barouch~\cite{wu-mc-tr-ba-76} to be expressible as an inf\/inite
sum of {\em holomorphic} functions,
given as multiple
integrals, denoted $ \chi^{(n)}$, that is
$ kT   \chi = \sum \chi^{(n)}$.
B.~Nickel found~\cite{nickel1, nickel2}
that each of these $ \chi^{(n)}$'s
is actually singular on a set of points located on the unit circle
$ |s| = |\sinh(2 K)| = 1$, where
$K= J/kT$ is the usual Ising model temperature variable.

These singularities are located at solution points of the following equations
 \begin{gather}
\label{sols}
{{1} \over {w}} =
 2
\left(s + {{1} \over {s}}\right) =
u^k + {{1} \over {u^k}}
+ u^m + {{1} \over {u^m}},
\\ \hbox{with} \quad
\qquad u^{2 n+1} = 1, \qquad
-n \le m, k \le n.\nonumber
\end{gather}
\looseness=1 From now on, we will call these singularities of the ``Nickelian type'',
or simply ``Nickelian singularities''.
The accumulation of this inf\/inite set of singularities of the higher-particle
components of $\chi(s)$ on the unit circle $ |s| = 1$,
leads, in the absence of mutual cancelation, to
some consequences regarding the
non holonomic (non D-f\/inite) character of the suscepti\-bility,
possibly building a natural boundary for the total $ \chi(s)$.
However, it should be noted that
 new singularities, that are not of the ``Nickelian type'',
were discovered as singularities of the Fuchsian linear dif\/ferential equation
associated~\cite{ze-bo-ha-ma-04,ze-bo-ha-ma-05c,ze-bo-ha-ma-05} with
 $\chi^{(3)}$ and as singularities of $\chi^{(3)}$
{\em itself\/}~\cite{bo-ha-ma-ze-07}
{\em but seen as a function of} $ s$.
They correspond to the quadratic polynomial
$ 1 +3 w +4 w^2$ where $2 w= s/(1+s^2)$.
In contrast with this situation, the Fuchsian linear dif\/ferential equation,
associated~\cite{ze-bo-ha-ma-05b} with $ \chi^{(4)}$, does not provide any
new singula\-rities.

\looseness=1
Some remarkable ``Russian-doll'' structure, as well as direct sum
decompositions, were found for the corresponding linear dif\/ferential
operators for $ \chi^{(3)}$ and $ \chi^{(4)}$.
 In order to understand the ``true nature''
of the susceptibility of the square lattice Ising model, it is of fun\-damen\-tal
importance to have a better understanding of the singularity structure
of the $n$-particle contributions~$\chi^{(n)}$, and also of
the {\em mathematical structures} associated with these
$ \chi^{(n)}$, namely the {\em infinite set} of (probably Fuchsian)
linear dif\/ferential equations associated with this inf\/inite set of holonomic
functions.
Finding more Fuchsian linear dif\/ferential equations having the $\chi^{(n)}$'s
as solutions, beyond those already found~\cite{ze-bo-ha-ma-04, ze-bo-ha-ma-05b}
for $ \chi^{(3)}$ and $ \chi^{(4)}$, probably requires the per\-formance
of a~large set of analytical, mathematical and
computer programming ``tours-de-force''.

 As an alternative, and in order to bypass this ``temporary''
obstruction, we have developed, in parallel, a new strategy.
We have introduced~\cite{bo-ha-ma-ze-07}
some single (or multiple) ``model'' integrals
as an ``ersatz'' for the $ \chi^{(n)}$'s as far as the locus
of the singularities is concerned.
The $ \chi^{(n)}$'s are def\/ined by
 $(n-1)$-dimensional integrals~\cite{nickel2,pt,yamada}
 (omitting the prefactor\footnote{The
prefactor reads $ (1-s^4)^{1/4}/s$ for
$ T > T_c$ and $ (1-s^{-4})^{1/4}$
for $T < T_c$ and in terms of the $w$ variable.})
\begin{gather}
\label{chi3tild}
\tilde{\chi}^{(n)} = {\frac{(2 w)^n}{n!}}
 \prod_{j=1}^{n-1}\int_0^{2\pi} {\frac{d\phi_j}{2\pi}}
\left( \prod_{j=1}^{n} y_j \right)   R^{(n)}
  \bigl( G^{(n)} \bigr)^2,
\end{gather}
where
\begin{gather}
\label{Gn}
G^{(n)} =
\left( \prod_{j=1}^{n} x_j \right)^{(n-1)/2}
 \prod_{1\le i < j \le n}
{\frac{2\sin{((\phi_i-\phi_j)/2)}}{1-x_ix_j}},
\qquad
R^{(n)} =
 {\frac{1+\prod\limits_{i=1}^{n}x_i}{1-\prod\limits_{i=1}^{n}x_i}} \nonumber
\end{gather}
with
\begin{gather}
\label{thex}
x_{i} = \frac{2w}{1-2w\cos (\phi _{i})
+\sqrt{\left( 1-2w\cos (\phi_{i})\right)^{2}-4w^{2}}}, \\
y_{i} =
\frac{1}{\sqrt{\left(1 -2 w\cos (\phi _{i})\right)^{2} -4w^{2}}},
\qquad \sum_{j=1}^n \phi_j= 0.\nonumber
\end{gather}

The two families of integrals we have
 considered in~\cite{bo-ha-ma-ze-07}
are very rough approximations of the integrals (\ref{chi3tild}).
For the f\/irst family\footnote{Denoted $ Y^{(n)}(w)$
 in~\cite{bo-ha-ma-ze-07}.}, we considered the $ n$-fold
integrals corresponding to the
 product of (the square\footnote{Surprisingly the integrand with
$ ( \prod_{j=1}^{n} y_j )^2$ yields
second order linear dif\/ferential equations~\cite{bo-ha-ma-ze-07}, and
 consequently, we have been able
to totally decipher the corresponding singularity structure.
By way of contrast the integrand with the simple product
$ ( \prod_{j=1}^{n} y_j )$ yields
 linear dif\/ferential equations of higher order, but with
identical singularities~\cite{bo-ha-ma-ze-07}.}
 of the) $y_i$'s,
integrated over the whole domain of integration of the
$\phi_i$ (thus getting rid of the factors
$ G^{(n)}$ and $ R^{(n)}$). Here, we found a
 subset of singularities occurring in
the $\chi^{(n)}$ {\em as well as the quadratic
polynomial condition} $ 1 +3w +4w^2 = 0$.

For the second family, we discarded the factor
$ G^{(n)}$ and the product of $ y_i$'s, and we restricted the domain of
integration to the principal diagonal of the angles $\phi_i$
($\phi_1 = \phi_2 = \cdots = \phi_{n-1}$).
These simple integrals (over a {\em single} variable), were
denoted~\cite{bo-ha-ma-ze-07} $ \Phi_{D}^{(n)}${\samepage
\begin{gather}
\label{chinaked}
\Phi_{D}^{(n)} =
 -{{1} \over {n!}} + {{2} \over {n!}} \int_0^{2\pi}
{\frac{d\phi}{2\pi}}
 {\frac{1}{1 -x^{n-1}(\phi)  x ((n-1)\phi)}},
\end{gather}
where $ x(\phi)$ is given by (\ref{thex}).}

Remarkably these very simple integrals both {\em reproduce
all the singularities}, discussed by
Nickel~\cite{nickel1,nickel2}, {\em as well as
the quadratic roots of} $ 1 +3w +4w^2 = 0$
found~\cite{ze-bo-ha-ma-04,ze-bo-ha-ma-05}
 for the linear ODE of $ \chi^{(3)}$.
One should however note that, in contrast with the
$ \chi^{(n)}$, no Russian-doll,
or direct sum decomposition structure, is found for the
linear dif\/ferential operators corresponding
to these simpler integrals $ \Phi_{D}^{(n)}$.

We return to the integrals (\ref{chi3tild})
where, this time, the natural next step is
to consider the following family of $ n$-fold integrals
\begin{gather}
\label{In}
\Phi_H^{(n)} = {\frac{1}{n!}}
 \prod_{j=1}^{n-1} \int_0^{2\pi} {\frac{d\phi_j}{2\pi}}
\left( \prod_{j=1}^{n} y_j \right)
 {\frac{1 +\prod\limits_{i=1}^{n} x_i}{1 -\prod\limits_{i=1}^{n} x_i}}
\end{gather}
which amounts to getting rid of the (fermionic) factor $ (G^{(n)})^2$
 in the $ n$-fold integral (\ref{chi3tild}).
 This family is as close as possible to
(\ref{chi3tild}), for which we know that
f\/inding the corresponding linear dif\/ferential ODE's
is a huge task.
The idea here is that the methods and techniques we have
developed~\cite{ze-bo-ha-ma-04,ze-bo-ha-ma-05}
for series expansions calculations of $ \chi^{(3)}$
and $ \chi^{(4)}$, seem to indicate that the
quite involved fermionic term $ (G^{(n)})^2$
in the integrand of (\ref{chi3tild})
should not impact ``too much'' on the location of singularities
 of these $ n$-fold integrals (\ref{chi3tild}).
This is the best simplif\/ication of the integrand of (\ref{chi3tild})
for which we can expect to retain much exact information about the
location of the singularities of the original Ising problem.
However, we certainly do not expect to recover from
the $ n$-fold integrals (\ref{In})
the local singular behavior (exponents,
amplitudes of singularities, etc \dots).
Getting rid of the (fermionic) factor $ (G^{(n)})^2$
are we moving away from the
elliptic curves of the two-dimensional Ising model?
Could it be possible that we lose the strong
(Russian-doll, direct sum decomposition)
algebro-dif\/ferential structures of the
 corresponding linear dif\/ferential operators inherited from
the second order dif\/ferential operator $ L_E$ (resp. $ L_K$)
of the complete elliptic integrals $ E$ (resp. $ K$),
but keep some characterization of elliptic curves
through more ``primitive'' (universal) features of these
$ n$-fold integral like the location of their singularities?

In the sequel, we give the expressions of $ \Phi_H^{(1)}$,
 $ \Phi_H^{(2)}$ and the Fuchsian
linear dif\/ferential equations for
$ \Phi_H^{(n)}$ for $ n = 3$ and $ n = 4$.
For $ n = 5, 6$, the computation
(linear ODE search of a series) becomes
much harder. Consequently we use a
{\em modulo prime} method to obtain the form of the
corresponding linear ODE with totally explicit singularity structure.
These results provide a large set of ``candidate singularities'' for
the $ \chi^{(n)}$.
From the resolution of the Landau
conditions~\cite{bo-ha-ma-ze-07,Smatrix} for (\ref{In}),
we have shown that the singularities of (the linear ODEs of)
these multiple integrals actually reduce to the concatenation
of the singularities of (the linear ODEs of)
 a set of one-dimensional integrals.
We discuss the {\em mathematical, as well as physical,
interpretation} of these new singularities.
In particular we can see that they correspond to {\em pinched
Landau-like singularities} as previously
noticed by Nickel~\cite{nickel-05}.
Among all these polynomial singularities, the
 quadratic numbers $ 1 +3 w +4 w^2 = 0$ are
highly selected. We will show that these selected
quadratic numbers are related to
{\em complex multiplication for the elliptic
curves} parameterizing the square Ising model.

 We present the multidimensional integrals $ \Phi_H^{(n)}$ and
the singularities of the corresponding linear
 ODE for $ n= 3, \dots, 6 $,
that we compare with the singularities
obtained from the Landau conditions. We have shown~\cite{bo-ha-ma-ze-07b} that the
set of singularities associated with the ODEs
of the multiple integrals $ \Phi^{(n)}_H$ reduce to the singularities of
the ODEs associated with a {\em finite number of one-dimensional integrals}.
Section~\ref{bridge} deals with the {\em complex
 multiplication for the elliptic
curves} related to the singularities given
 by the zeros of the quadratic polynomial
$ 1 +3w +4 w^2$.

\section[The singularities of the linear ODE for $ \Phi_H^{(n)}$]{The singularities of the linear ODE for $\boldsymbol{\Phi_H^{(n)}}$}
\label{singODE}

For the f\/irst two values of $n$, one obtains
\begin{gather*}
\Phi_H^{(1)} =  {\frac{1}{1-4w}}
\qquad \mbox{and}\qquad
\Phi_H^{(2)} = {{1} \over {2}}   {\frac{1}{1-16w^2}}
{_2}F_1 \big( 1/2, -1/2; 1; 16w^2\big).
\end{gather*}

For $n \ge 3$, the series coef\/f\/icients of the multiple
integrals $ \Phi_H^{(n)}$ are obtained by expanding in the variables
$x_i$ and performing the integration (see Appendix A of~\cite{bo-ha-ma-ze-07b}).
One obtains{\samepage
\begin{gather}
\label{integratedsum}
\Phi_H^{(n)} =
 {\frac{1}{n!}}   \sum_{k=0}^{\infty} \sum_{p=0}^{\infty}
(2 -\delta_{k,0})   (2 -\delta_{p,0})   w^{n(k+p)}   a^n(k,p),
\end{gather}
where $ a(k,p)$ is a $ _4 F_3$ hypergeometric
 series dependent on $ w$.}

The advantage of using these simplif\/ied integrals
 (\ref{In}) instead of the original
ones (\ref{chi3tild}) is twofold.
Using (\ref{integratedsum}) the series generation
 is straightforward compared to
the complexity related to the $\chi^{(n)}$.
As an illustration note that on a desk
computer, $ \Phi_H^{(n)}$ are generated up to $ w^{200}$
 in less than 10 seconds CPU time
for all values of $ n$,
while the simplest case of the $ \chi^{(n)}$, namely~$ \chi^{(3)}$,
 took three minutes to generate the series up to $ w^{200}$.
This dif\/ference between
the $ \Phi_H^{(n)}$ and $ \chi^{(n)}$ increases rapidly with
increasing $ n$ and increasing number of generated terms.
We note that for the
$ \Phi_H^{(n)}$ quantities and for a f\/ixed order, the CPU time
is decreasing\footnote{This can be seen from the
series expansion (\ref{integratedsum}).
Denoting $ R_0$ the f\/ixed order, one has
 $ n   (p+k) \le R_0$, while
the CPU time for the series generation of $ a^n(k, p)$
 is not strongly dependent on $ n$.} with
 increasing~$ n$. For~$ \chi^{(n)}$
the opposite is the case.
The second point is that,
for a given $n$, the linear ODE can be found with less terms in the series
compared to the linear ODE for the $\chi^{(n)}$. Indeed for $\chi^{(3)}$,
360 terms were needed while 150
terms were enough for $ \Phi_H^{(3)}$. The same feature holds
for $\chi^{(4)}$ and $ \Phi_H^{(4)}$ (185 terms for $\chi^{(4)}$
and 56 terms\footnote{From now on,
 for even $ n$, the number of terms stands
for the number of terms in the variable $x=w^2$.}
 for $ \Phi_H^{(4)}$).

With the fully integrated sum (\ref{integratedsum}),
 a suf\/f\/icient number of terms is generated to obtain the linear
 dif\/ferential equations.
We succeeded in obtaining the linear dif\/ferential equations, respectively
of minimal order f\/ive and six, corresponding to
$ \Phi_H^{(3)}$ and $ \Phi_H^{(4)}$. These linear ODE's
 are given in Appendix \ref{a}.

For $ \Phi_H^{(n)}$ ($n \ge 5$), the calculations, in order to get
the linear ODEs become really huge\footnote{Except
the generation of long series which remains reasonable.}.
For this reason, we introduce a modular strategy which
amounts to generating long series {\em modulo a prime} and
then deducing the ODE modulo that prime.
Note that the ODE of minimal order is {\em not necessarily
the simplest one} as far as the required number of terms
in the series expansion to f\/ind the linear ODE is concerned.
We have already encountered such a situation~\cite{ze-bo-ha-ma-05b,Diag}.
For $ \Phi_H^{(5)}$ (resp. $ \Phi_H^{(6)}$), the
 linear ODE of minimal order is
of order 17 (resp.~27) and needs 8471 (resp.~9272)
terms in the series expansion to be found.

Actually, for $ \Phi_H^{(5)}$ (resp.~$ \Phi_H^{(6)}$), we have found
the corresponding linear ODEs of order 28 (resp.~42) with {\em only}
2208 (resp.~1838) terms from which we have deduced the
minimal ones.
The form of these two minimal order linear ODEs obtained
modulo primes is sketched in Appendix~\ref{a}.
In particular, the singularities (given by the roots
 of the head polynomial in front of the highest
order derivative), are given with the corresponding multiplicity in Appendix~\ref{a}.
Some details about the linear ODE search are also given in Appendix~\ref{a}.

We have also obtained very long series (40000 coef\/f\/icients) modulo
primes for $ \Phi_H^{(7)}$, but, unfortunately, this has not
been suf\/f\/icient to identify the linear ODE (mod.~prime) up to order~100.

The singularities of the linear ODE for the f\/irst $ \Phi_H^{(n)}$
 are respectively zeros of the following polynomials (besides $w=\infty$):
\begin{alignat}{3}
& n=3, \quad  && w  \left( 1-16 w^2 \right)
 \left( 1-w \right)
\left( 1+2 w \right) \left( 1+3 w+4 {w}^{2} \right), & \nonumber \\
& n=4,\quad   && w   \left( 1-16 w^2 \right)
 \left(1-4 w^2 \right), & \nonumber \\
& n=5, \quad   && w   \left( 1-16 w^2 \right)
 \left( 1-w^2 \right)
\left( 1+2 w \right) \left( 1+3 w+4 {w}^{2} \right)
\left( 1-3 w+{w}^{2} \right) & \nonumber \\
&&& \quad {} \times\left( 1+2 w-4 {w}^{2}\right) \left( 1+4 w+8 {w}^{2} \right)
 \left( 1-7 w+5 {w}^{2}-4 {w}^{3} \right) & \nonumber \\
&& & \quad{}\times
 \left( 1-w-3 {w}^{2}+4 {w}^{3} \right)
 \left( 1+8 w+20 {w}^{2}+15 {w}^{3}+4 {w}^{4} \right),\label{singphi5} & \\
& n=6, \quad   && w   \left( 1-16 w^2 \right)
 \left( 1-4 {w}^{2} \right)
 \left( 1-{w}^{2} \right)
 \left( 1-25 {w}^{2} \right)
 \left( 1 +3w +4w^2 \right) & \nonumber \\
& && \quad{}\times \left( 1-9 {w}^{2} \right)
 \left(1-3w+4w^2\right)
 \left( 1-10 {w}^{2}+29 {w}^{4} \right).\label{singphi6}
\end{alignat}

For $ n= 7$ and $ n= 8$, besides modulo primes
 series calculations mentioned above,
we also generated very long series from which we obtained
in f\/loating point form, the polynomials given
in Appendix \ref{singphi7phi8} (using generalised dif\/ferential Pad\'e methods).

If we compare the singularities for the ODEs for the $ \Phi_H^{(n)}$ to those obtained with
the ``Diagonal model''\footnote{Not to be confused with the
``diagonal susceptibility'' and the corresponding~\cite{Diag} $ n$-fold
integrals $ \chi^{(n)}_d$.} presented
in~\cite{bo-ha-ma-ze-07}, i.e.\ for the ODEs for the $\Phi_D^{(n)}$,
one sees that the singularities of the linear ODE for the ``Diagonal model''
are identical to those of the linear ODE of the $ \Phi_H^{(n)}$
 for $n= 3, 4$ (and are a proper subset to those of
 $ \Phi_H^{(n)}$ for $n = 5, 6$).
The additional singularities for $n= 5, 6$ are zeros of the polynomials:
\begin{alignat*}{3}
& n=5, \quad   && \left( 1+3 w+4 {w}^{2} \right)
 \left( 1+4 w+8 {w}^{2} \right)
\left( 1-7 w+5 {w}^{2}-4 {w}^{3} \right), \nonumber & \\
& n=6, \quad  && \left( 1 +3w +4w^2 \right) \left(1-3w+4w^2\right)
 \left( 1-25 {w}^{2} \right). \nonumber&
\end{alignat*}
For $ n = 7$, the zeros of the following polynomials (among others)
are singularities which are not
of Nickel's type (\ref{sols}) and do not occur
for $\Phi_D^{(n)}$:
\begin{gather*}
  1 +8w +15w^2 -21w^3 -60 w^4
+16 w^5 +96 w^6 +64w^7, \nonumber \\
 1 -4w -16w^2 -48w^3
 +32 w^4 -128 w^5. \nonumber
\end{gather*}

The linear ODEs of the multiple integrals $ \Phi_H^{(n)}$ thus
display {\em additional singularities}
 for $n= 5, 6$ and $n=7$ ($n= 8$
see below) compared to the
linear ODE of the single integrals $ \Phi_D^{(n)}$.

We found it remarkable that the linear ODEs
for the integrals $\Phi_D^{(n)}$ display all the ``Nickelian singularities'' (\ref{sols}) ,
as well as the new quadratic numbers $1+3w+4w^2= 0$ found for $\chi^{(3)}$.
It is thus interesting to see how the singularities for $ \Phi_D^{(n)}$
are included in the singularities for $\Phi_H^{(n)}$ and whether
the new (with respect to $\Phi_D^{(n)}$) singularities can be given
by one-dimensional integrals similar to $\Phi_D^{(n)}$.
Let us mention that the singularities of
 the linear ODE for $ \Phi_H^{(3)}$
(resp.~$ \Phi_H^{(4)}$) {\em are remarkably also singularities}
 of the linear ODE for
$ \Phi_H^{(5)}$ (resp.~$ \Phi_H^{(6)}$).
In~\cite{bo-ha-ma-ze-07b} it was shown how this comes about and how it generalizes.
For this, we had to solve the Landau conditions~\cite{bo-ha-ma-ze-07b} for
the $n$-fold integrals~(\ref{In}).

\section{Bridging physics and mathematics}
\label{bridge}

In a set of papers~\cite{Fuchs,PainleveFuchs} and
 in the previous sections, we have underlined
the central role played by the {\em elliptic parametrization}
of the Ising model, in particular
 the role played by the second order linear
dif\/ferential operator $ L_E$ (or~$ L_K$) corresponding
 to the complete elliptic integral~$ E$ (or~$ K$),
and the occurrence of an inf\/inite
number of {\em modular curves}~\cite{Holo},
canonically associated with {\em elliptic curves}.
We are getting close to identify the
lattice Ising model, (or more generally Baxter model),
with the {\em theory of elliptic curves}. In such an identif\/ication
framework one may seek for
``special values'' of the modulus $ k$
 that could have a~``physical meaning'',
as well as a~``mathematical interpretation'' (beyond just being singularities),
as singularities of the $\chi^{(j)}$.

\subsection{Revisiting the theory of elliptic curves with a physics viewpoint}
\label{revisit}

The deep link between the theory of elliptic curves and the theory of modular
forms is now well established~\cite{rhoades}. More
 simply the crucial role of the
modular group in analysing elliptic curves is well known. For that reason
seeking ``special values~\cite{selected}'' of the modulus $ k$,
that might have a ``physical meaning''
as well as a mathematical meaning, as singularities of the $\chi^{(j)}$,
it may be interesting to, alternatively, introduce
 the {\em modular function} called the
$ j$-function which corresponds to Klein's
absolute invariant multiplied by
$ (12)^3 = 1728$
\begin{gather}
\label{jfun}
j(k) = 256
 {\frac { \left( 1-{k}^{2}+{k}^{4} \right)^{3}}{ k^{4}
 \left( 1-k^{2} \right)^{2}}}
\end{gather}
and, alternatively, seek for ``special values''
 of the $ j$-function (\ref{jfun}),
since it automatically takes into account
 the {\em modular symmetry group}
of the problem.
The modular group requires one to introduce
the period ratio and the {\em nome}
of the elliptic functions.
The elliptic nome, def\/ined in terms
of the periods of the elliptic functions, reads
\begin{gather}
\label{defq}
q = \exp \left(-\pi {\frac{K(1-k^2)}{K(k^2)}} \right)
 = \exp(i \pi \tau),
\end{gather}
where $\tau$ is the half period ratio\footnote{In the theory of
modular forms $q^2$ is also used instead of $ q$.
In number theoretical literature the half-period ratio is
taken as $-i \tau$.}.

The $ SL(2, Z)$ transformations of the modular group
\begin{gather*}
\tau \  \longrightarrow \
 {{ a \tau + b} \over { c \tau +d}}
\end{gather*}
which preserve the $ j$-function (\ref{jfun}),
should not be confused with {\em isogenies}
of elliptic curves like the {\em Gauss or Landen} transformations
\begin{gather}
\label{gausslanden}
 \tau \rightarrow 2 \tau, \qquad
 \tau \rightarrow {\frac{\tau}{2}},
\end{gather}
and, more generally ($n$ integer)
\begin{gather}
\label{ntau}
 \tau \rightarrow n  \tau, \qquad
 \tau \rightarrow {\frac{\tau}{n}}.
\end{gather}
which actually modify the $ j$-function (\ref{jfun}),
but are ``compatible'' with the lattice of
periods (the inclusion of one lattice into the other one).

Roughly speaking, and as far as the elliptic curves
of the Ising model (resp. Baxter model)
are concerned, the $ SL(2, Z)$
transformations of the modular group
are invariance symmetries (reparametrizations), while the
transformations (\ref{ntau}) are highly non-trivial
covariants that we will see as {\em exact representations of
the renormalization group}.

\subsection{Landen and Gauss transformations
as generators\\ of the exact renormalization group}
\label{renorm}

Let us consider the complete elliptic integral $ K(k)$ def\/ined as:
\begin{gather*}
K(k) =
 {_2}F_1 \bigl( 1/2, 1/2; 1; k \bigr).
\end{gather*}
Two relations between
$K(k)$, evaluated at two dif\/ferent modulus,
can be found in, e.g.~\cite{erdeleyi} and
read\footnote{Note that
the relation (\ref{relat1}) is valid for any value of the
modulus $k$, while the validity of
(\ref{relat2}) is restricted to $\vert k \vert <1$.}:
\begin{gather}
\label{relat1}
(1+\sqrt{1-k^2})  K\left( k^2 \right) = 2
K \left( {\frac{(1-\sqrt{1-k^2})^2}{(1+\sqrt{1-k^2})^2}} \right),
\\
\label{relat2}
(1+k)  K\left( k^2 \right) =
K \left( {\frac{4 k}{(1+k)^2}} \right).
\end{gather}
The arguments in $K$, on the right-hand-side
 of (\ref{relat1}), (\ref{relat2}), are the square of the modulus
$k$ transformed by the so-called (descending)
Landen or (ascending) Landen
(or Gauss) transformations:
\begin{gather}
\label{landenD}
 k \  \longrightarrow \  k_{-1} =
 {\frac{1-\sqrt{1-k^2}}{1+\sqrt{1-k^2}}},
\\
\label{landenA}
 k \  \longrightarrow \  k_{1} =
{{ 2 \sqrt{k}} \over {1+k}}.
\end{gather}
A sequence of such transformations can be used to evaluate (numerically),
in a rapidly convergent way, the elliptic integrals
from iterations of~(\ref{landenD}) or of~(\ref{landenA}). Changing $k$
 to the complementary modulus
 $ k'= \sqrt{1-k^2}$, and likewise for
the transformed $k$,
the half period ratio transforms through
(\ref{landenD}), (\ref{landenA}), like (\ref{gausslanden}).

The {\em real} f\/ixed points of the transformations
 (\ref{landenD}) and (\ref{landenA})
are $ k = 0$
 (the trivial inf\/inite or zero temperature points) and
$ k = 1$ (the ferromagnetic and antiferromagnetic
{\em critical point} of the square Ising model).
Iterating (\ref{landenD}) or (\ref{landenA}), one
converges, respectively to
$ k = 0$ or $ k = 1$.
In terms of the half period ratio,
this reads, respectively, $\tau=\infty$ and $\tau=0$
which {\em correspond to a degeneration
 of the elliptic parametrization
into a rational parametrization}.
In view of these f\/ixed points, it is natural to identify the
transformations (\ref{landenD}) or (\ref{landenA}), and, more generally,
any transformation\footnote{See relation (1.3) in~\cite{Schwartz}.}
$ \tau \rightarrow n   \tau$
or $ \tau \rightarrow \tau/n$ ($n$ integer), as
{\em exact generators of the renormalization group}.
It is a straightforward exercise, using
the identities (\ref{relat1}), (\ref{relat2}),
 to write a ``renormalization recursion''
on the internal energy $ U$ of the Ising model
\begin{gather*}
U(k) =
 {{c} \over {2 s}}
 +
 {{c   (c^2-2)} \over {2 s}}   K(k^2),
\end{gather*}
where $ c$ and $ s$ denote $ \cosh(2K)$ and $ \sinh(2K)$
respectively.

\subsection[Complex multiplication of elliptic curves
and fixed points of Landen transformations]{Complex multiplication of elliptic curves\\
and f\/ixed points of Landen transformations}
\label{complex}

Since we are interested in singularities {\em in the complex plane}
of some ``well-suited'' variable~($s$,~$ k$, $ w$),
one should {\em not} restrict
 (\ref{landenD}) and (\ref{landenA}) as transformations
 on {\em real variables}, restricting
to {\em real fixed points} of these transformations,
but actually consider the f\/ixed points of these transformations
seen as transformations on {\em complex variables}.

For instance, if one considers (\ref{landenA}) as an
algebraic transformation
 of the {\em complex variable} $ k$ and
solve $ k_1^2 -k^2= 0$, one obtains:
\begin{gather*}
k   (1-k)  ({k}^{2} +3k +4) = 0.
\end{gather*}
The roots of
\begin{gather}
\label{polk}
k^2 + 3 k +4 = 0,
\end{gather}
are (up to a sign)
 {\em fixed points} of (\ref{landenA}).
We thus see the occurrence of {\em additional non-trivial complex
selected values} of the modulus $ k$, beyond the
well-known values $k= 1, 0, \infty $, corresponding to
degeneration of the elliptic curve into a {\em rational curve},
and physically, to the {\em critical} Ising model
 and to (high-low temperature)
trivializations of the model.

Of course, when extending (\ref{landenA})
 to {\em complex values}, one can be concerned about
 keeping track of the sign of $ k_{1}$ in (\ref{landenA})
in front of the square root $ \sqrt{k}$.
Reference \cite{bo-ha-ma-ze-07b} provides a similar f\/ixed point calculation
for (\ref{landenA}) extended to complex values, but for a representation
of (\ref{landenA}) in term of the
 modular $ j$-function. Such calculations
single out the remarkable integer value $ j = (-15)^3$, which
 is known to be one of the {\em nine Heegner numbers} (see \cite{bo-ha-ma-ze-07b}).
It is important to note that this representation
of (\ref{landenA}) in term of the modular $ j$-function
is the well-known fundamental modular curve
{\em symmetric} in $ j$ and $ j_1$
 (see \cite{bo-ha-ma-ze-07b,Hanna,Myers})
\begin{gather*}
 j^2 j_1^2 -(j+ j_1)
(j^2 + 1487 j j_1 + j_1^2)
-12 \cdot 30^6 (j + j_1) \nonumber \\
\qquad {}+ 3 \cdot 15^3  (16 j^2
 - 4027 j j_1
+ 16 j_1^2) +8 \cdot 30^9
= 0.
\end{gather*}
which represents,
{\em at the same time}, the Landen and Gauss
transformations (\ref{gausslanden})
as a consequence of the modular invariance ($\tau \leftrightarrow 1/\tau$).

A straightforward calculation of the elliptic nome (\ref{defq}) gives,
for the polynomial (\ref{polk}) and the polynomial
deduced from the Kramers--Wannier duality $k \rightarrow 1/k$ respectively,
exact values for $\tau$, the half period
ratio, as very simple {\em quadratic numbers}:
\begin{gather*}
\tau_1 = {{ \pm 3 +i \sqrt{7} } \over {4}},
\qquad
\tau_2 = {{\pm 1 +i \sqrt{7} } \over {2}}.
\end{gather*}
These quadratic numbers correspond to
{\em complex multiplication} and to
$j = (-15)^3$. These two quadratic numbers
are such that $2 \tau_1 \mp 1 =\tau_2$.
Let us focus on $\tau_2$ for which we can we write
\begin{gather*}
\tau = 1 -{{2} \over {\tau}}.
\end{gather*}

Taking into account the two modular group involutions
 $\tau \rightarrow 1 -\tau$
and $ \tau \rightarrow 1/\tau$,
we f\/ind that $1 - 2/\tau$ is, {\em up to the modular group},
equivalent to $ \tau /2$.
The quadratic relation $ \tau^2 - \tau +2 = 0 $ thus
amounts to looking at the f\/ixed
 points of the Landen transformation
$ \tau \rightarrow 2 \tau$ {\em up to the modular group}.
 This is, in fact a quite general statement: the
{\em complex multiplication}
values can all be seen as f\/ixed points, {\em up to the modular group},
of the generalizations of Landen transformation,
namely $ \tau \rightarrow n \tau$
for $ n$ integer (here $\simeq$ denotes
the equivalence up to the modular group):
\begin{gather*}
 \tau^2 - \tau +n = 0
\qquad  \hbox{or} \qquad
 \tau = 1 -{{n} \over {\tau}}
 \simeq n  \tau, \qquad
\tau = {{ 1 + i \sqrt{4 n-1}} \over {2}}.
\end{gather*}
Complex multiplication corresponds to integer
values of the modular $ j$-function
(as in the case of the Heegner numbers see \cite{bo-ha-ma-ze-07b}).

For elliptic curves in f\/ield of characteristic zero,
the only well-known selected set
 of values for $k $ corresponds
(besides $k = 0, 1, \infty$) to the
values for which the elliptic curve has {\em complex
multiplication}~\cite{broglie}, and we see these selected values, here,
{\em as fixed points, in the complex plane,} of transformations
(isogenies)
that are {\em exact representations of generators of the
renormalization group}.

It is now totally natural to see if the singularities
we have obtained for the $ n$-fold integrals~(\ref{In}),
{\em can be interpreted in the framework of elliptic curve theory},
in terms of this physi\-cally, and mathematically, highly selected
set of values for elliptic curves, namely complex
multiplication values.

\subsection[Complex multiplication for
$ 1+3 w + 4 w^2 = 0$]{Complex multiplication for
$\boldsymbol{1+3 w + 4 w^2 = 0}$}
\label{1plus3w}

Let us consider the f\/irst unexpected singularities
$ 1+3 w + 4 w^2 = 0$
we found~\cite{ze-bo-ha-ma-04,ze-bo-ha-ma-05}
 for the Fuchsian linear dif\/ferential
equation of $ \chi^{(3)}$,
and also found in other $n$-fold integrals of the Ising
class~\cite{bo-ha-ma-ze-07}.
This polynomial condition reads in the $ s$ variable,
$ \left( 2 {s}^{2}+s+1 \right)
 \left( {s}^{2}+s+2 \right) = 0$.
We have shown~\cite{bo-ha-ma-ze-07} that
$ \chi^{(3)}$ {\em itself} is not singular
at the roots of the f\/irst polynomial whose roots are such that
$\vert s \vert <1$, but is actually singular at the roots
 of the second polynomial. In the variable $k= s^2$,
 these singularities read:
\begin{gather}
\label{polk2}
(4 k^2 +3 k +1 ) (k^2 +3 k +4) = 0.
\end{gather}
The second polynomial has actually been seen
 to correspond to {\em fixed points}
of the Landen transformation (see (\ref{polk})).
Note that the two
polynomials in (\ref{polk2}) are related by the Kramers--Wannier
 duality $ k \rightarrow 1/k$ (and therefore both correspond
 to the same value of the modular $ j$-function: $ j = (-15)^3$).

In other words we see that the selected
values $ 1+3 w + 4 w^2 = 0$,
occurring in the (high-temperature) susceptibility of the Ising model
as singularities of the three-particle term $ \chi^{(3)}$,
actually correspond to the {\em occurrence of complex
multiplication on the elliptic curves} of the Ising model,
and can also be seen {\em as fixed points
of the renormalization group
when extended to complex values of the modulus} $ k$.

Let us note that the occurrence of Heegner numbers and {\em complex
multiplication} has already occurred in
other contexts, even if the statement
was not explicit. In the framework of
 the construction of Liouville f\/ield theory,
Gervais and Neveu suggested~\cite{Gervais}
new classes of critical statistical
 models (see Appendix \ref{ee}), where, besides
 the well-known $N$-th root of unity situation,
they found the following selected values
of the {\em multiplicative crossing}~\cite{Rammal2} $ t$:
\begin{gather}
\label{othervaluesint}
 t = e^{ i \pi (1+i \sqrt{3})/2}
 = i  e^{-\pi \sqrt{3}/2 }, \\
\label{othervaluesint2}
 t = e^{ i \pi (1+i)}
 = -e^{-\pi}.
\end{gather}
 If one wants to see this
multiplicative
crossing~\cite{multiplicative3,multiplicative2,multiplicative1,multiplicative} as
 a modular nome (see \cite{bo-ha-ma-ze-07b}),
the two previous situations actually correspond to
selected values of the modular $ j$-function namely
$ j((1+i \sqrt{3})/2) = (0)^3$ for (\ref{othervaluesint}),
and $ j(1+i) = (12)^3$ for (\ref{othervaluesint2}), which
actually correspond to {\em Heegner numbers
and complex multiplication}~\cite{broglie}.
 It is however important not to
feed the confusion already too prevalent
 in the literature, between
 a {\em ``temperature-like'' nome} like (\ref{defq})
and a {\em multiplicative
crossing modular nome} (see Appendix \ref{ee}).
In the Baxter model~\cite{Baxter}, the f\/irst is denoted by $ q$ and the
second one by $ x$.
In fact one probably has, {\em not one, but two modular groups} taking place,
one acting on the ``temperature-like'' nome $ q$
 and the other one acting on the multiplicative
crossing $ x$. We will not go further along this quite speculative line
which amounts to introducing {\em elliptic quantum groups}~\cite{Mano}
and (see Appendix \ref{ee}) {\em elliptic gamma functions} (generalization
 of theta functions\footnote{The partition function of the Baxter model
can be seen as a ratio and product of elliptic
 gamma functions and theta functions,
it is thus naturally expressed as a double inf\/inite
product. Similar double, and even triple, products
 appear in correlation functions of
 the eight vertex model~\cite{JKKMW,JMN}. })
 which can be seen~\cite{Varchenko} as ``automorphic forms of degree 1'',
 when the Jacobi modular forms are ``automorphic forms of degree 0''
and are associated (up to simple semi-direct products)
to $ SL(3, Z)$ instead of $ SL(2, Z)$.

\subsection[Beyond $ 1 +3 w + 4 w^2 = 0$]{Beyond $\boldsymbol{1 +3 w + 4 w^2 = 0}$}
\label{beyo}

As a consequence of the fact that the modular $ j$-function is a
function of $ w^2 $, the quadratic polynomial condition
$ 1 -3 w + 4 w^2 = 0$,
corresponds to the same selected values of the
 modular $ j$-function as $ 1 +3 w + 4 w^2 = 0$,
 namely $ j = (-15)^3$.
 The quadratic polynomial $ 1 -3 w + 4 w^2 = 0$
actually occurs in the singularities of
the linear ODE for $ \Phi^{(6)}_H$
(and all the higher $ \Phi^{(2n)}_H$, if one believes formulas
(28) and (29) in~\cite{bo-ha-ma-ze-07b}).

In view of the remarkable mathematical (and physical)
 interpretation of the quadratic
values $ 1+3 w + 4 w^2 = 0$,
(and also $ 1-3 w + 4 w^2 = 0$)
in terms of {\em complex multiplication, or fixed points of the
renormalization group}, it is natural to see
if such a ``complex multiplication'' interpretation
also exists for other singularities of $ \chi^{(n)}$, and
as a f\/irst step, for the singularities of the linear dif\/ferential equations
of our $ n$-fold integrals (\ref{In}), that
we expect to identify, or at least,
 have some overlap with the singularities of the $ \chi^{(n)}$.

We have found two other polynomial conditions
which correspond to remarkable integer values of the modular $ j$-function.
The singularities $ 1 -8 w^2= 0 $ correspond to
$j=(12)^3$ and $\tau=\pm 1 +i$ (see \cite{bo-ha-ma-ze-07b}). They correspond to
``Nickelian singularities'' (\ref{sols})
 for $\chi^{(8)}$ (and thus~$\Phi^{(8)}_H$)
and to ``non-Nickelian singularities'' for $\Phi_D^{(10)}$ and
$\Phi_H^{(10)}$.
Another polynomial condition is $ 1 -32 w^2= 0 $, which give
``non-Nickelian singularities'' that begin to appear at $n=10$.
These singularities correspond to the integer value of
the modular $ j$-function, $j=(66)^3$ and
to $\tau=2 i$ and $\tau =-4/5+i 2/5$.

Among the singularities of the linear ODE for $\Phi_H^{(n)}$ given in
(\ref{singphi5}), (\ref{singphi6}) or obtained from the formula (29)
given in \cite{bo-ha-ma-ze-07b} up to $n=15$, we have found no other
singularity identif\/ied with the remarkable Heegner numbers~\cite{jFunc}
or, more generally, with other selected values of the
 modular $ j$-function, associated to complex multiplication.

Could it be that the (non-Nickelian) singularities
(\ref{singphi5}), (\ref{singphi6}), which
 do not match with complex multiplication
of the elliptic curves, are actually selected for
mathematical structures more complex or more general
 than elliptic curves (possibly
 linked~\cite{Varchenko} to $ SL(3, Z)$
 instead of $ SL(2, Z)$ modular group)?
This could amount to moving away from the isotropic
Ising model towards the Baxter model. At f\/irst
sight the analysis of the anisotropic Ising
 model~\cite{Fuchs} could be considered
as a f\/irst step in that ``Baxter-model''
direction.
The selected situations for elliptic functions and complete elliptic
integrals, would thus, be generalized to the search of
``selected situations'' of their multidimensional generalizations
(Lauricella, Kamp\'e de Feri\'e, Appell, \dots) {\em that we have actually seen
to occur in the anisotropic Ising model}~\cite{Fuchs} and even in our
series expansions of $ \chi^{(3)}$ and $ \chi^{(4)}$.

Along similar lines,
one may recall the $ n$-fold integrals introduced by
Beukers, Vasilyev~\cite{Va2,Va1} and Sorokin~\cite{So2,So3}
\begin{gather*}
\int_{0}^{1} \cdots \int_{0}^{1}
 dt_1 dt_2 \cdots dt_{a+1}
\prod_{j=1}^{a+1} {{ (1-t_j) t_j^r }
 \over { (z -t_1 t_2 \cdots t_{a+1})^3 }}
\end{gather*}
and other well-poised hypergeometric functions
\begin{gather*}
 {{((2 r +1) n +2)! } \over {n!^{2 r+1}}}
 z^{(r+1) n + 1}
 \int_{0}^{1} \cdots \int_{0}^{1}
 dt_1 dt_2 \cdots dt_{a+1} \\
 \qquad {} \times
 \left( {{ \prod\limits_{j=1}^{a+1} (1-t_j) t_j^r}
\over { (z -t_1 t_2 \cdots t_{a+1})^{2 r+1}}} \right)^n
 {{ z +t_1 t_2 \cdots t_{a+1} }
 \over { (z -t_1 t_2 \cdots t_{a+1})^3 }}, \nonumber
\end{gather*}
or the Goncharov--Manin integrals~\cite{Fischler} which occur
in the {\em moduli space of curves}~\cite{Bloch,Goncharov}.
These integrals~\cite{Brown,Fischler3,Fischler2,Huttner,Huttner2,Krattenthaler,Racinet}
 look almost the same as the ones
we have introduced and analyzed in the study of the
diagonal susceptibility of the Ising model~\cite{Diag}.

It is worthy to recall that $ \zeta(3)$ appeared
in some of our ``connection matrix method'' results for
the dif\/ferential Galois group~\cite{ze-bo-ha-ma-05c}
of the Fuchsian linear ODE for $ \chi^{(3)}$ and $ \chi^{(4)}$,
and the occurrence of zeta functions in
many $ n$-fold integrals.
Also recall that Feynman amplitudes can be seen as periods in
 the ``motivic sense''~\cite{Bloch},
 and are often linked to multiple zeta numbers.
Along this line, the following integral~\cite{Fischler3,Fischler}
deals with $ \zeta(3)$:
\begin{gather}
\label{otherzeta3}
I_n (z) =
 \int_{0}^{1} du dv dw
 {{ (1-u)^n u^n   (1-v)^n v^n   (1-w)^n w^n }
 \over { (1 -u v)^{n+1}   (z -u v w)^{n+1} }}.
\end{gather}

From the series expansion of this holonomic $ n$-fold integral,
we have obtained~\cite{bo-ha-ma-ze-07b} an order four Fuchsian linear
dif\/ferential equation (see Appendix \ref{fin}).
On such linear dif\/ferential operators the ``logarithmic''
nature of these integrals becomes clear. The occurrence of linear
dif\/ferential operators is not a complete surprise
if one recalls that in Ap\'ery's proof of the irrationality
of~$ \zeta(3)$ a crucial role is played
 by the linear dif\/ferential operator~\cite{Beukers}
\begin{gather*}
 L = (t^2 -34 t +1) t^2   D_t^3
+ (6 t^2 -153 t +3) t   D_t^2 + (7 t^2 -112 t +1)   D_t
 + (t-5)
\end{gather*}
this operator being linked to the modularity of the algebraic variety:
\begin{gather*}
 x + {{ 1} \over {x}} + y + {{ 1} \over {y}}
+ z + {{ 1} \over {z}} + w + {{ 1} \over {w} }
 = 0.
\end{gather*}

These $ n$-fold integrals have to be compared with the
 (more involved) $ n$-fold Ising integrals
corresponding to the $ \chi^{(n)}$, and
to the {\em theory of elliptic curves} (rather than
 rational curves $ CP_1$ in the previously cited
examples~\cite{Brown,Fischler4,Fischler3,Fischler2,Huttner,Huttner2,Krattenthaler,Racinet}),
we try to underline in this paper.

With these new singularities,
are we exploring some remarkable ``selected
situations'' of some moduli space of curves~\cite{Adler,Pierre}
corresponding to pointed~\cite{Bloch2,Bloch4,Bloch3} (marked)
curves~\cite{Barad}, instead
of simple elliptic curves~\cite{Harris}?
In practice this will probably just correspond to considering
a {\em pro\-duct}
 of $ n$ times a rational or elliptic curve
 minus some sets of remarkable
algebraic varieties~\cite{Diag}, $ x_i x_j = 1$,
 $ x_i x_j x_k = 1$,
hyperplanes $ x_ i = x_j$, $\dots$.

\section{Conclusion}
\label{conclu}

We have displayed several examples of $ n$-fold {\em holonomic} integrals
associated with the two-dimensional Ising model on a square lattice~\cite{crandall}.
The corresponding
linear dif\/ferential operators with polynomial coef\/f\/icients are shown to be very
closely linked to the {\em theory of elliptic curves} (and modular forms) and display
many remarkable structures (Russian-doll structure,
direct sum structure, complex multiplication
as selected singular values for these operators, \dots).
These linear dif\/ferential operators are not only Fuchsian operators, they are
Fuchsian operators with {\em rational exponents}: the various indicial
polynomials corresponding to all the regular singularities of these linear
 dif\/ferential operators
have only {\em rational} (or integer)
roots. It is tempting to try to understand these deep algebraico-dif\/ferential structures
as a consequence of the {\em underlying elliptic curve in the Ising model}, or
more generally, of some algebraic varieties built from this elliptic curve
({\em product of curves}, \dots), or corresponding to the integrands
of these $ n$-fold integrals.
Could it be possible that these large number of remarkable properties
have a {\em geometrical interpretation} (generalisation
of hypergeometric functions and Picard--Fuchs systems, Grif\/f\/iths
 cohomology of hypersurface of $ CP_n$, rigid
 local systems~\cite{Bertin,Gerkmann,Katz11,Katz2,Katz3,Katz0,Katz1,pic}, \dots)
 with a strong background of {\em algebraic geometry}?
One could, for instance, imagine
that these various $ n$-fold {\em holonomic} integrals
might be interpreted as {\em periods
of some algebraic varieties}, all the strong and
 deep algebraico-dif\/ferential structures
we have displayed in this paper, being a
consequence of this very {\em rigid geometrical framework}.
The central role played by the {\em theory of elliptic curves}
and their isomonodromic deformations (Painlev\'e equations) for the Ising model
on a {\em lattice} is also underlined in the fundamental f\/inite-dif\/ference
 (non-linear overdetermined) system of quadratic functional
relations~\cite{Perk3,mcc-wu-80,Perk2,PerkDunk}
 (see (\ref{perk}) in Appendix \ref{painl}) for
the two-point correlation functions
of the Ising model on the square lattice. As Painlev\'e
and (discrete) integrability
specialists call it, these {\em lattice
equations} are f\/inite-dif\/ference generalisation of Painlev\'e
equations and they have a lot of very deep
consequences : they are, for instance, the very reason why the
susceptibility series can be calculated
from a program with {\em polynomial growth}~\cite{or-ni-gu-pe-01b}. Such
 an {\em overdetermined} system (\ref{perk}) 
can be seen as generating
an inf\/inite number of non-trivial identities on the complete
elliptic integrals of the f\/irst and second kind.

It is important to note that all these remarkable structures and deep symmetries
(remarkable functional identities,
algebraico-dif\/ferential structures, modular forms,
continuous~\cite{mazz,okamoto} and discrete Painlev\'e structures, \dots),
underline the central role played by the
{\em theory of elliptic curves} for the two-dimensional
Ising model {\em on a lattice}.
Note that a large part of these remarkable
 structures and deep ({\em lattice}) symmetries
is {\em lost in the scaling limit}. In the
scaling limit some of these remarkable structures
remain (the Russian-doll telescopic embedding of the
 linear dif\/ferential operators), but, for instance,
the direct-sum structure is lost. The scaling limit
 yields the occurrence of an {\em irregular
singularity at infinity}: the Fuchsian character
of the linear operators is lost, as well as most
of the {\em remarkable structures associated with the underlying elliptic curve theo\-ry}.
For instance for two-point correlation functions,
the complete elliptic integrals of f\/irst and second kind
 $ K$ and $ E$ are replaced by {\em modified
Bessel functions} (with their irregular
singularity at inf\/inity), but the fact that
form factors are simple polynomial expressions
of $ E$ and $ K$ is lost: the form factors, in the scaling limit, {\em are not
 simple polynomial expressions of modified
 Bessel functions}. In the scaling limit,
a large part of the strong background of
algebraic geometry that exists {\em on the lattice
model}, and yields so many remarkable deep and strong
 structures and symmetries, seems to disappear.
If the geometrical interpretation we suggested for the lattice model exist, could it be
possible that it is essentially lost in the
 scaling limit, the underlying algebraic varieties
necessary for this geometrical interpretation being lost, or becoming some complicated
analytical manifolds? Recalling the emergence of an {\em irregu\-lar singularity}
(at inf\/inity), an irregu\-lar singularity can,
in principle, be understood~\cite{Garnier} as a {\em confluence of two regular
 singularities} (for complete elliptic integrals of f\/irst and second kind
 we have the conf\/luence of the two regular singularities $ 0$, $ \infty$
among the three regular singularities $ 0$, $ 1$, $ \infty$). To our knowledge
we have not often seen\footnote{With the exception of C.~Zhang
explaining~\cite{Zhang} Ramis' conf\/luent approach~\cite{Ramis3,Ramis,Ramis2}
 of irregular singularities.}
 in the litterature the structures associated to irregular
singularities (Stokes multipliers, singular behaviours, \dots)
be obtained as a ``conf\/luent limit'' of the structures associated with the two regular
singularities. From a general viewpoint, in a desire
to see analytical manifolds as a conf\/luent limit of algebraic varieties,
 one can imagine that the structures of
the Ising model in the scaling limit could, in principle, be
 obtained from a~(very involved) ``conf\/luent limit''
of the remarkable structures deeply linked to the theory of elliptic curves
that exist for the Ising model
{\em specifically on the lattice}. This remains to be done.
In practice we see that, paradoxically
from a criticality-universality mainstream viewpoint, the (of\/f-critical, non-universal)
Ising model {\em on a lattice has much deeper, and fundamental,
structures than the Ising model in the scaling limit}.
Note that the results of holonomic and algebraic-geometry
nature we have displayed in this paper, are {\em not} specif\/ic of the two-dimensional
Ising model, or, even, of free-fermion models. We have
{\em not used the free-fermion character
of the Ising model}. We have heavily used the {\em elliptic parametrisation}
of the two-dimensional lattice Ising model. One can imagine that many of these results,
and structures, exist for Yang--Baxter integrable models with an elliptic
parametrisation (the Baxter model~\cite{Korepin2,Boos2006,Korepin1}, \dots),
and, more generally, for any Yang--Baxter integrable model\footnote{It is not even clear
that one has to restrict oneself to Yang--Baxter integrability: $ n$-fold
integrals associated with particle physics, polyzeta functions,
Feynman diagrams, etc \dots seem to indicate that one has a much more general framework
for these ideas~\cite{Kreimer2000,Levin}.}, the central role of the
elliptic curve being replaced by the relevance of the algebraic variety emerging in the
 Yang--Baxter equations (higher genus curves~\cite{Hypergeometric}, Abelian
varieties, \dots). Integrable models on a lattice are probably deeper, and dressed
with much more symmetries and remarkable structures\footnote{See for instance
Baxter's concept of Z-invariance~\cite{Perk5,Z}.}, than their
scaling limits.
Such an apparently paradoxical (for the
 f\/ield theory mainstream) conclusion is certainly not a~surprise
for Painlev\'e and (discrete) integrability
 specialists who are used to see, and understand,
{\em lattice equations as deeper, and more
fundamental}~\cite{Sakai1,Sakai2}, than the dif\/ferential equations.

\appendix

\section[Quadratic partial difference Painlev\'e generalisations]{Quadratic partial dif\/ference Painlev\'e generalisations}
\label{painl}

Quadratic partial dif\/ference equations were
shown~\cite{Perk3,mcc-wu-80,Perk2,PerkDunk} to be
satisf\/ied by two-point correlation functions of the two-dimensional
Ising model on the square lattice. These quadratic partial dif\/ference
equations (valid in the anisotropic case),
 are actually valid for the $ \lambda$-extension
of the two-point correlation functions
 $ C(M, N; \lambda)$
{\em for any value of} $ \lambda$ :
\begin{gather}
s^2 [C(M, N; \lambda)^2
 -C(M, N-1; \lambda) C(M, N+1; \lambda)]
 \nonumber \\
\qquad{} + [ C^{*}(M, N; \lambda)^2
-C^{*}(M-1, N; \lambda) C^{*}(M+1, N; \lambda)]
 = 0, \label{perk}\\
 {s'}^2  [C(M, N; \lambda)^2
 -C(M-1, N; \lambda) C(M+1, N; \lambda)] \nonumber \\
 \qquad {}+ [ C^{*}(M, N; \lambda)^2
 -C^{*}(M, N-1; \lambda) C^{*}(M, N+1; \lambda)]
 = 0, \nonumber \\
  s {s'}   [C(M, N; \lambda) C(M+1, N+1; \lambda)
 -C(M, N+1; \lambda)
 C(M+1, N; \lambda)] \nonumber \\
 \qquad  = C^{*}(M, N; \lambda)
 C^{*}(M+1, N+1; \lambda) -C^{*}(M, N+1; \lambda)
 C^{*}(M+1, N; \lambda)] \nonumber
\end{gather}
with $ s = \sinh(2 K)$, $ s = \sinh(2 K)$,
where $ K$ and $ K'$ correspond to the horizontal and vertical
coupling constant of the anisotropic square Ising model
and where $ C^{*}$ are the $ \lambda$-extension
of the dual correlation functions.

\section[Algebraic solutions of PVI for
 $\lambda=\cos(\pi m/n)$ and modular curves]{Algebraic solutions of PVI for
 $\boldsymbol{\lambda=\cos(\pi m/n)}$\\ and modular curves}
\label{aa}

The unexpectedly simple expressions
for the form factors $ f^{(j)}_{N,N}$ of
Sections \ref{new}--\ref{scal}, and the corresponding
remarkable dif\/ferential structures,
may be used to obtain many further results.
We displayed some of these results in Section~6 of \cite{Holo}.
Recalling that, when $\lambda= 1$, the Ising
correlation functions $ C(N,N; 1)$ satisfy
Fuchsian linear dif\/ferential equations~\cite{PainleveFuchs}
with an order that
grows with $ N$, it is quite natural to inquire
 whether there are any other values of $ \lambda$
for which $ C(N,N;\lambda)$ will satisfy a
Fuchsian linear dif\/ferential equation. One such
family of $ \lambda$ is motivated
 by the work of Cecotti and Vafa~\cite{cv} on
$ N =2$ supersymmetric f\/ield theories where they encountered
$\lambda$ extensions of the Ising
correlations in the scaling limit~\cite{mtw}
with ($m $ and $ n$ are integers):
\begin{gather*}
\lambda = \cos(\pi m/n).
\end{gather*}
Indeed, we have found that for
$ n = 3,\dots, 20$, the functions $ C(N,N;\lambda)$
satisfy Fuchsian linear dif\/ferential equations
 whose orders, in contrast with those of the
$ \lambda= 1$ equations~\cite{PainleveFuchs},
 {\em do not} depend on $ N$.

The function $ C(N,N;\lambda)$ is such that its log-derivative
is actually a solution of the sigma form of Painlev\'e VI : it is
a {\em transcendent}
 function ``par excellence''. However, the unexpectedly simple
expressions for these form factors $ f^{(j)}_{N,N}$,
 strongly suggest to try to
resum the inf\/inite sums~(\ref{formm1}), and~(\ref{formp1}),
of form factors $ f^{(j)}_{N,N}$, corresponding
to the function $ C(N,N;\lambda)$,
and see if these transcendent functions could be ``less complex'' than
one can imagine
at f\/irst sight, at least for a set of ``singled-out''
 values of $ \lambda$. For instance,
 are there any values of $\lambda \neq 1$
which share, with $ \lambda =1$, the
property that $ C(N,N;\lambda)$ satisf\/ies
a Fuchsian linear dif\/ferential equation?

Actually, introducing, instead of the modulus $ k$
of elliptic functions (for $ T > T_c$, $ k = s^2$),
or the $ s$ and $ t$ variables,
 the {\em nome} of the elliptic functions
 (see relations (5.7)--(5.11) in~\cite{or-ni-gu-pe-01b}),
we have been able to perform such a resummation,
 getting, {\em for arbitrary} $ \lambda$, nice closed
expressions for the $ C(N,N;\lambda)$ for the f\/irst
values of $ N$, ($N = 0, 1 , 2, \dots $),
as sums of ratios of theta functions (and their derivatives), corresponding
 to {\em Eisenstein series}, or {\em quasi-modular
forms}. These results will be displayed
 in forthcoming publications. The simplest
example corresponds to $ N = 0$ where
 $ C_{-}(N,N;\lambda)$
is just the ratio of two
Jacobi $ \theta_3$ functions
\begin{gather}
\label{closed}
C_{-}(0,0;\lambda) =
{{ \theta_3(u,q)} \over {\theta_3(0,q)}},
 \qquad  \hbox{where}
\qquad\lambda = \cos(u).
\end{gather}
All these results strongly suggest to focus
 on $ u = \pi m/n$ ($m$ and $ n$ integers)
yielding for the possible choice of ``selected'' values of $ \lambda$:
\begin{gather}
\label{special}
\lambda = \cos(\pi m/n).
\end{gather}
Actually these special values (\ref{special}) of
$ \lambda $ already occurred in a study of $ N=2$
 supersymmetric f\/ield theories~\cite{cv}
in a similar series construction of solutions of the Painlev{\'e} V
(or Painlev{\'e}~III for a ratio of functions)
equation for the scaling limit of the Ising model~\cite{wu-mc-tr-ba-76}.

Recalling the quadratic f\/inite dif\/ference equations~\cite{Perk,mcc-wu-80}
 (\ref{perk})
we can deduce that the of\/f-diagonal terms $ C(M,N;\lambda)$
are, in the isotropic case, algebraic expressions of sum of ratios of theta
 functions and their derivatives. For
 the singled-out values $ \lambda = \cos(\pi m/n)$
 ($ n$ and $ m$ integers), the of\/f-diagonal terms $ C(M,N;\lambda)$
are, in the isotropic case, algebraic expressions of the variable~$ t$: do these
algebraic expressions also correspond to modular curves? Actually
they clearly single out
$ t = 0, 1, \infty$, \dots.

\medskip

\noindent
{\bf Remark.} For this set of selected values of $ \lambda$
the $ \lambda$-extension $ C(N, N; \lambda)$ are seen to
 be algebraic expressions of the variable $ t$
and, more remarkably, associated with a {\em modular curve}
 $ P(C, t) = 0$ ($P$ denotes a polynomial with integer
coef\/f\/icients, $ C$ denotes $ C(N, N; \lambda)$
for $ \lambda = \cos(\pi m/n)$, for certain
 integer values of $ n$ and $ m$,
and the only branching points are for $ t = 0, 1, \infty$).
The fact that the only singular points are $ t = 0, 1, \infty$
 can be seen to be inherited
from the fact that the $ \lambda$-extension $ C(N, N; \lambda)$
 is actually solution of (\ref{jimbo-miwa})
{\em for any} $ \lambda$: the sigma form of Painlev\'e VI,
namely (\ref{jimbo-miwa}), naturally singles out
 $ t = 0, 1, \infty$
(and only $ t = 0, 1, \infty$).

\section[Linear differential equations of some $\Phi_H^{(n)}$]{Linear dif\/ferential equations of some $\boldsymbol{\Phi_H^{(n)}}$}
\label{a}

\subsection[Linear ODE for $ \Phi_H^{(3)}$]{Linear ODE for $\boldsymbol{\Phi_H^{(3)}}$}

The minimal order linear dif\/ferential equation satisf\/ied by
$\Phi_H^{(3)}$ reads
\begin{gather*}
\sum_{n=0}^{5} a_{n}(w)
 {\frac{{d^{n}}}{{dw^{n}}}}F(w) = 0,
 \end{gather*}
 where
\begin{gather*}
a_5 (w) =
\left(1-w \right) \left(1-4 w \right)^{4}
 \left( 1+4 w \right)^{2}
\left( 1+2 w \right) (1+3 w+4 {w}^{2})  {w}^{3}
 P_5(w), \nonumber \\
a_4 (w) =
\left(1-4 w \right)^{3} \left( 1+4 w \right)  {w}^{2}
  P_4(w), \qquad a_3 (w) =
-2 \left(1-4 w \right)^{2}  w P_3(w), \nonumber \\
a_2 (w) = 2 (1-4 w)  P_2(w), \qquad
a_1 (w) = -8 P_1(w), \qquad
a_0 (w) = -96 P_0(w), \nonumber
\end{gather*}
and where the apparent singularities
 polynomial $ P_5(w)$ in the head polynomial reads
\begin{gather*}
 P_5(w) =
-5+21 w+428 {w}^{2}+5364 {w}^{3}-82416 {w}^{4}
-299504 {w}^{5}+714944 {w}^{6}
 \nonumber \\
\phantom{P_5(w) =}{}  +3127872 {w}^{7}-8220672 {w}^{8} -25858048 {w}^{9}-7077888 {w}^{10}
 +31424512 {w}^{11}
\nonumber \\
\phantom{P_5(w) =}{} -42467328 {w}^{12}
-31457280 {w}^{13}-4194304 {w}^{14}+4194304 {w}^{15},
 \nonumber
\end{gather*}
the other polynomials $ P_n(w)$ are given in~\cite{bo-ha-ma-ze-07b}.

\subsection[Linear ODE for $ \Phi_H^{(4)}$]{Linear ODE for $\boldsymbol{\Phi_H^{(4)}}$}

The minimal order linear dif\/ferential equation satisf\/ied by
$\Phi_H^{(4)}$ reads (with $x=16w^2$)
\begin{gather*}
\sum_{n=0}^{6} a_{n}(x)  {\frac{{d^{n}}}{{dx^{n}}}}F(x)
 = 0,
\end{gather*}
where
\begin{gather*}
 a_6 (x) = 64 \left(x-4 \right)
 (1-x)^{4}{x}^{4}  P_6(x),\qquad
 a_5 (x) =
 -128 (1-x)^{3} {x}^{3}   P_5(x),
\nonumber \\
 a_4 (x) = 16 (1-x)^{2}{x}^{2}
  P_4(x), \qquad a_3 (x) = -64 (1-x) x  P_3(x),
\nonumber \\
a_2 (x) = -4   P_2(x), \qquad
 a_1 (x) = -8  P_1(x), \qquad
a_0 (x) = -3 (1-x)   P_0(x), \nonumber
\end{gather*}
where the apparent singularities polynomial $ P_6(x)$
 in the head polynomial reads:
\begin{gather*}
P_6(x) =
 128+2233 x-2847 {x}^{2} +3143 {x}^{3}
-3601 {x}^{4}+144 {x}^{5}-64 {x}^{6}. \nonumber
\end{gather*}
The other polynomials $ P_n(w)$ are given in~\cite{bo-ha-ma-ze-07b}.

\subsection[Linear ODE modulo a prime for  $ \Phi_H^{(5)}$]{Linear ODE modulo a prime for  $\boldsymbol{\Phi_H^{(5)}}$}
\label{subsec}

The linear dif\/ferential equation of minimal order seventeen
satisf\/ied by
$\Phi_H^{(5)}$ is of the form
\begin{gather*}
\sum_{n=0}^{17} a_{n}(w)
{\frac{{d^{n}}}{{dw^{n}}}}F(w) = 0,
\end{gather*}
with
\begin{gather*}
 a_{17} (w) =
 \left( 1-4 w \right)^{12} \left( 1+4 w \right)^{9}
( 1-w)^{2}
\left( w+1 \right) \left( 1+2 w \right)\left( 1+3 w+4 {w}^{2} \right)^{2}
\nonumber \\
\phantom{a_{17} (w) =}{}  \times
\left( 1-3 w+{w}^{2} \right)
\left( 1+2 w-4 {w}^{2} \right) (1+4 w+8 {w}^{2})\left( 1-7 w+5 {w}^{2}-4 {w}^{3} \right)\nonumber \\
\phantom{a_{17} (w) =}{} \times
  \left( 1-w-3 {w}^{2}+4 {w}^{3} \right)
\left( 1+8 w+20 {w}^{2}+15 {w}^{3}+4 {w}^{4} \right)  {w}^{12}
 P_{17}(w), \nonumber \\
a_{16} (w) =
{w}^{11}  \left( 1-4 w \right)^{11} \left( 1+4 w \right)^{8}
( 1-w)
\left( 1+3 w+4 {w}^{2} \right)
 P_{16}(w), \nonumber \\
a_{15} (w) =
{w}^{10}  \left( 1-4 w \right)^{10} \left( 1+4 w \right)^{7}
 P_{15}(w), \nonumber \\
a_{14} (w) =
{w}^{9} \left( 1-4 w \right)^{9} \left( 1+4 w \right)^{6}
 P_{14}(w), \nonumber \\
\cdots \cdots \cdots \cdots \cdots \cdots \cdots \cdots \cdots \cdots \cdots \cdots \cdots  \nonumber
\end{gather*}
where the $ 430$ roots of $ P_{17}(w)$ are {\em apparent singularities}.
The degrees of these polynomials $ P_{n}(w)$
 are such that the degrees of $a_i(w)$ are decreasing as:
$ \deg(a_{i+1}(w))= \deg(a_i(w)) +1$.
In fact, with 2208 terms we have found the ODE of
$ \Phi_H^{((5)}$ at order $ q= 28$ using the following ansatz for
the linear ODE search ($Dw$ denotes $d/dw$)
\begin{gather*}
\sum_{i=0}^{q} z(i) p(i)  Dw^i
\end{gather*}
with
\begin{gather*}
z(i) = w^{\alpha(-1 + i)}
 (1 - 16 w^2)^{\alpha(-1 + i)}  z_0^{\alpha(1 + i- q)} \nonumber
\end{gather*}
where $ \alpha(n) = \min(0, n) $ and:
\begin{gather*}
 z_0 = (1 + w) (1 - w) (1 + 2 w)
 (1 - 3 w + w^2)
(1 +2 w - 4 w^2) (1 + 3 w + 4 w^2) \nonumber \\
\phantom{z_0 =}{}  \times (1 + 4 w + 8 w^2)
(1 - 7 w + 5 w^2- 4 w^3)(1 - w - 3 w^2 + 4 w^3) \nonumber \\
\phantom{z_0 =}{} \times
 (1 + 8 w + 20 w^2 + 15 w^3 +4 w^4)
\nonumber
\end{gather*}
the $ p(i)$ being the unknown polynomials. The minimal
 order ODE is deduced from the set of
linearly independent ODEs found at order~28. Instead
 of these linear ODEs with quite large apparent singularities
polynomials, we can provide an alternative linear ODE of higher order
with {\em no apparent singularities}.
This is the so-called ``desingularization'' procedure of a linear
ODE. The price to pay to get rid of the large apparent
polynomial can be that the higher order ODE with
no apparent polynomial may not be Fuchsian anymore
(because of an irregular singularity at inf\/inity). One can
also consider desingularizations {\em preserving Fuchsianity}.

We give below the linear dif\/ferential operator of order~25,
modulo the prime $ 27449$. At this order, the
linear dif\/ferential operator has {\em no apparent singularities}
\begin{gather*}
\left( w+12523 \right) ^{2} \left( w+3989 \right) \left( w+9241 \right)
 \left( w+21789 \right) ^{2} \left( w+20587 \right)^{9} \nonumber \\
 \qquad {} \times
\left( {w}^{3}+6861 {w}^{2}+6864 w+6862 \right) \left( w+6862 \right)^{12}
 \left( w+27448 \right)^{2}\nonumber \\
 \qquad {}\times \left( {w}^{3}+14565 {w}^{2}
+6775 w+7627 \right) \left( w+13725 \right) \nonumber \\
\qquad {}\times
 \left( {w}^{3}+20586 {w}^{2}+6862 w+20587 \right)
 \left( w+4483 \right)
 \left( w+8965 \right) \nonumber \\
\qquad {}\times \left( w+19750 \right) \left( w+18481 \right)
 \left( w+1 \right) \left( w+9736 \right)   {w}^{12}   Dw^{25}
 \nonumber \\
\qquad{}+  1134 \left( w+12523 \right)   P_{23} \left( w+21789 \right)
\left( w+20587 \right)^{8} \nonumber \\
\qquad {} \times
\left( w+6862 \right)^{11} \left( w+27448 \right) {w}^{11}
 {{Dw}}^{24}\nonumber \\
\qquad{} + 24190 \left( {w}^{2}+26576 w+1920 \right) \left( w+6798 \right)
 P_{17} \left( w+20587 \right)^{7}\nonumber \\
 \qquad {}\times
 \left( w+6862 \right)^{10}   P_{10}  P_{9}
  Q_{9} {w}^{10}  {{ Dw}}^{23}
 \nonumber \\
\qquad{}+  18886   P_{7} \left( w+16985 \right)
 \left( w+20587 \right)^{6}
\left( w+6862 \right)^{9}  P_{22}
\nonumber \\
 \qquad{} \times \left( {w}^{4}+1460 {w}^{3}
+5534 {w}^{2}+16322 w+22102 \right)
 Q_{7} P_{11} P_{51} {w}^{9}  {{Dw}}^{22}
 \nonumber \\
\qquad {}+2514 {w}^{8} P_{82}
 \left( {w}^{2}+24026 w+4328 \right)
 \left( w+20587 \right)^{5}  P_{14}
\nonumber \\
\qquad{} \times \left( {w}^{6}+834 {w}^{5}+1985 {w}^{4}
+9290 {w}^{3}+16354 {w}^{2}+4746 w+4283 \right) \nonumber \\
\qquad{} \times \left( w+6862 \right)^{8}
 \left( w+10153 \right)  {{ Dw}}^{21}
\nonumber \\
\qquad{} +7703  P_{24} {w}^{7} \left( {w}^{2}+21361 w+14160 \right)
 \left( w+20587 \right)^{4}   P_{80} \nonumber \\
\qquad {} \times \left( w+6862 \right)^{7}
 \left( w+12463 \right)  {{\it Dw}}^{20}
\nonumber \\
\qquad{}+7921   P_{14} {w}^{6} \left( {w}^{2}+9437 w+10364 \right)
 P_{7} \left( w+20587 \right)^{3}   P_{62}
 \nonumber \\
\qquad {}\times \left(w+4208 \right)   P_{10}
 \left( w+3899 \right)   {{Dw}}^{19}
\nonumber \\
\qquad{}+26675 \left( {w}^{2}+7849 w+652 \right)   P_{61}
 \left( w+20587 \right)^{2} \left( w+6862 \right)^{5}
P_{42}
 \nonumber \\
\qquad {}\times
 \left( {w}^{6}+16220 {w}^{5}+14100 {w}^{4}+16063 {w}^{3}
+18759 {w}^{2}+27385 w+22000 \right) \nonumber \\
\qquad {} \times \left( w+6862 \right)^{6}
 P_{12} {w}^{5}  {{Dw}}^{18}
 \nonumber \\
\qquad{} +6905 {w}^{4}  P_{9} \left( w+20587 \right)
 \left( w+6862 \right)^{4}  Q_{9} P_{71} \left( w+21055 \right)
 \left( w+5262 \right)
\nonumber \\
\qquad {}\times
 \left( {w}^{5}+24655 {w}^{4}+5244 {w}^{3}
+12152 {w}^{2}+16121 w+18032 \right)
 P_{17}  {{Dw}}^{17}
\nonumber \\
\qquad{}+12508 \left( w+10977 \right) {w}^{3} \left( w+6862 \right)^{3}
 \left( {w}^{3}+8559 {w}^{2}+11289 w+9639 \right)\nonumber \\
\qquad {}\times
 \left( {w}^{3}+14200 {w}^{2}+21840 w+15970 \right) \\
 \qquad{}\times
 \left( {w}^{5}+2056 {w}^{4}+12162 {w}^{3}+6126 {w}^{2}+22729 w
+16548 \right) \nonumber \\
\qquad {}\times P_{16}  P_{77}
\left( {w}^{2}+16577 w+6876 \right)
  P_{8}   {{Dw}}^{16}
\\
\qquad{}+8317 {w}^{2} \left( w+21290 \right) \left( {w}^{2}+4503 w+23586 \right)
 \nonumber \\
\qquad {}\times
\left( {w}^{5}+26400 {w}^{4}+26455 {w}^{3}
+17876 {w}^{2}+7063 w+15487 \right)
\nonumber \\
\qquad {}\times \left({w}^{6}+10703 {w}^{5}+1763 {w}^{4}+12902 {w}^{3}
+6843 {w}^{2}+2123 w+18674 \right)\nonumber \\
\qquad {}\times
P_{43}  P_{45} \left( w+17888 \right)
\left( w+6862 \right)^{2}   P_{13}  {{Dw}}^{15}
\nonumber \\
\qquad{}+ 13440  P_{68} w \left( w+6862 \right)   P_{48}
\left( w+13354 \right)   {{Dw}}^{14}
+11221  P_{27}  P_{39}  P_{8}
 Q_{22}  P_{22}  {{Dw}}^{13}
\nonumber \\
\qquad{}+4954 \left( w+15012 \right)   P_{46}
 \left( {w}^{2}+15293 w+14779 \right)
 P_{68}   {{Dw}}^{12}
\nonumber \\
\qquad{}+ 26649  P_{106}  P_{8} \left( w+11457 \right)
 \left( w+24683 \right)   {{Dw}}^{11}
\nonumber \\
\qquad {}+ 20358 \left( {w}^{2}+2503 w+22492 \right)   P_{22}
 \left( {w}^{3}+13156 {w}^{2}+2684 w+8942 \right)   P_{7}
 \nonumber \\
\qquad {}\times
 \left( {w}^{4}+15516 {w}^{3}+4945 {w}^{2}+4117 w+4952 \right)
  P_{77}   {{Dw}}^{10}
 \nonumber \\
\qquad{}+18965 \left( {w}^{2}+10905 w+26489 \right) \nonumber \\
\qquad {}\times
 \left( {w}^{5}+13584 {w}^{4}+17617 {w}^{3}
+21809 {w}^{2}+8787 w+16558 \right)
 P_{11}  P_{40}  P_{56}  {{Dw}}^{9}
\nonumber \\
\qquad{}+16238 \left( {w}^{5}+6787 {w}^{4}+23738 {w}^{3}
+20731 {w}^{2}+22246 w+21284 \right) \nonumber \\
 \qquad {}\times \left( {w}^{6}+13024 {w}^{5}+11465 {w}^{4}
+2372 {w}^{3}+6908 {w}^{2}+612 w+10791 \right) \\
\qquad{}\times{}  \left( w+12744 \right)   P_{101}  {{Dw}}^{8}
+ 706 \left( w+3771 \right)   P_{65}  P_{46}  {{Dw}}^{7}
\nonumber \\
\qquad {}+16695 \left( {w}^{5}+8888 {w}^{4}+6808 {w}^{3}
+26162 {w}^{2}+12699 w+21123 \right) \nonumber \\
\qquad {}\times
 \left( {w}^{3}+14297 {w}^{2}+5276 w+11625 \right)   P_{79}
\left( {w}^{2}+17252 w+11894 \right) \nonumber \\
\qquad {}\times
\left( {w}^{5}+1644 {w}^{4}+15551 {w}^{3}
+18064 {w}^{2}+8721 w+21521 \right)
 P_{17}  {{Dw}}^{6}
\nonumber \\
\qquad {} +19013  P_{24}  P_{31}
 \left( {w}^{3}+4946 {w}^{2}+25614 w+25774 \right)
\nonumber \\
\qquad {}\times
 \left( {w}^{5}+884 {w}^{4}+16517 {w}^{3}
+2210 {w}^{2}+13068 w+15270 \right)
  P_{37}  P_{10}  {{Dw}}^{5}
\nonumber \\
\qquad {} + 12971 \left( {w}^{2}+7070 w+2438 \right)
   P_{25} \nonumber \\
\qquad {}\times \left( {w}^{6}+869 {w}^{5}+26611 {w}^{4}+21387 {w}^{3}
+25750 {w}^{2}+21941 w+24186 \right) \nonumber \\
\qquad {}\times
 \left( w+23774 \right)   P_{17}
 P_{16}  P_{42}  {{Dw}}^{4}
\nonumber \\
\qquad{} +19579   P_{29}
 P_{11} \left( w+19783 \right) \left( w+9767 \right)
   P_{7}   P_{36}
 \left( w+15147 \right)   P_{22}   {{Dw}}^{3}
\nonumber \\
\qquad{}+9934 \left( {w}^{6}+7835 {w}^{5}+21707 {w}^{4}+5095 {w}^{3}
+8057 {w}^{2}+13773 w+14223 \right)
 \nonumber \\
\qquad{}\times
\left( {w}^{4}+8474 {w}^{3}+9011 {w}^{2}+12522 w+22902 \right)
P_{16}   P_{72}
 P_{9}   {{Dw}}^{2}
\nonumber \\
\qquad{}+1875   Q_{24}  P_{51}  P_{7}  P_{24}  {Dw}
+4564  \left( w+25617 \right)   P_{104},
\nonumber
\end{gather*}
where the $ P_{n}$ and $ Q_{n}$ is a ``short'' notations
to encounter polynomials of degree $ n$
(they can be dif\/ferent from one coef\/f\/icient of the $ m$-th derivative
$ Dw^m$ to another).

\medskip

\noindent
{\bf Remark.} We sketch such a quite tedious result (if we give explicitly the undef\/ined
polynomials the result would be really huge \dots) to give the reader some hint of
how such an exact result modulo a prime looks like: the exact expressions of the various polynomials,
which are the coef\/f\/icients in front of the derivatives, can actually
 be factorized modulo prime {\em without any ambiguity}.
For instance the factor $\left( w+27448 \right)^{2}$ in the head polynomial
(coef\/f\/icient of $ Dw^{25}$) is nothing but $\left( w-1 \right)^{2}$
modulo the prime $27449$. The interest of such an exact calculation is that we can {\em exactly} compare
the various factors in the head polynomial with a set of polynomials
 we have conjectured to be singularities of the linear ODE. We can totally conf\/irm the existence
 of some of (or all) these conjectured polynomials, and discriminate between apparent singularities
and ``true'' singularities.
The prime is large enough to {\em avoid any ambiguity corresponding to accidental factorisations}
(because the prime would be too small): this is conf\/irmed by the same calculations performed for
other similar large enough primes.

\subsection[Linear ODE modulo a prime for $ \Phi_H^{(6)}$]{Linear ODE modulo a prime for $\boldsymbol{\Phi_H^{(6)}}$}
\label{subsec6}

The linear dif\/ferential equation of {\em minimal order} (namely twenty-seven),
satisf\/ied by $\Phi_H^{(6)}$, reads (with $x=w^2$)
\begin{gather*}
\sum_{n=0}^{27} a_{n}(x)  {\frac{{d^{n}}}{{dx^{n}}}}F(x)
 = 0,
 \end{gather*}
 with
\begin{gather*}
a_{27} (x) =
\left( 1-16 x \right)^{16}
\left( 1-4 x \right)^{3}
\left( 1-x \right)
\left( 1-25 x \right)
\left( 1-9 x \right) {x}^{21}
\nonumber \\
 \phantom{a_{27} (x) =} {}\times \left( 1-x+16 {x}^{2} \right)
\left( 1-10 x+29 {x}^{2} \right)
  P_{27}(x),
\nonumber \\
a_{26} (x) =
\left( 1-16 x \right)^{15}
\left( 1-4 x \right)^{2} {x}^{20}
  P_{26}(x),
\nonumber \\
a_{25} (x) =
\left( 1-16 x \right)^{14}
\left( 1-4 x \right) {x}^{19}
  P_{25}(x),
\nonumber \\
a_{24} (x) =
\left( 1-16 x \right)^{13} {x}^{18}
  P_{24}(x),  \\
 \cdots \cdots \cdots \cdots \cdots \cdots \cdots  \cdots \cdots \cdots\nonumber
\end{gather*}
where the $307$ roots of $P_{27}(x)$ are {\em apparent singularities}.
The degrees of the $P_n(w)$ polynomials are such that
the degrees of $a_i(w)$ are decreasing as:
$\deg(a_{i+1}(w)) = \deg(a_i(w))+1$

In fact, with 1838 terms we have found the linear ODE of
$ \Phi_H^{(6)}$ at order $q= 42$
 using the following ansatz for
the linear ODE search ($Dx$ denotes $d/dx$)
\begin{gather*}
\sum_{i=0}^{q} z(i) p(i)   Dx^i
\end{gather*}
with
\begin{gather*}
z(i) = x^{\alpha(-1 + i)}
(1 - 16 x)^{\alpha(-1 + i)}   z_0^{\alpha(1 + i -q)},
\end{gather*}
where $ \alpha(n) = \min(0, n) $ and
\begin{gather*}
z_0 = (1 - 25 x) (1 - 9 x)
 (1 - 4 x) (1 - x)  (1 - x +16 x^2) (1 - 10 x + 29 x^2),
\end{gather*}
the $ p(i)$ being the unknown polynomials. The minimal order ODE is deduced from the set of
linearly independent ODEs found at order 42.

Here also, instead of this linear ODE with
quite a large apparent singularities
polynomial, we can provide an alternative linear ODE of higher order
with no apparent singularities (but it may not be Fuchsian anymore).
We give in the following the linear dif\/ferential operator,
modulo the prime $ 32749$, of order 30. At this order, the
linear dif\/ferential operator has {\em no apparent singularities}
\begin{gather*}
\left( x+8187 \right)^{3} \left( {x}^{2}+10234 x+22515 \right)
 \left( x+10234 \right)^{16} \left( x+32748 \right) {x}^{21} \nonumber \\
 \qquad {}\times
 \left( x+10113 \right) \left( x+31439 \right) \left( x+14555 \right)
 \left( x+15860 \right)   {{Dx}}^{30} \nonumber \\
\qquad{}+ 8594 \left( x+8187 \right)^{2} \left( x+10234 \right)^{15}
 \left( x+551 \right) {x}^{20} \nonumber \\
 \qquad {}\times \left( {x}^{4}+22258 {x}^{3}+24734 {x}^{2}
+32441 x+31408 \right)  P_{85}  P_{21}  {{Dx}}^{29}
\nonumber \\
\qquad{}+ 30840 \left( x+8187 \right) \left( x+10234 \right)^{14}
 \left( x+28552 \right)   P_{50}   P_{61} \nonumber \\
 \qquad {}\times \left( {x}^{3}+5377 {x}^{2}
+7622 x+28946 \right) {x}^{19}   {{Dx}}^{28}
\nonumber \\
\qquad{}+ 22635 \left( x+10234 \right)^{13}{x}^{18} \left( x+22434 \right)
 P_{16}  P_{23}  P_{69}  P_8   {{\it Dx}}^{27}
\nonumber \\
\qquad{} + 15369 \left( x+10234 \right)^{12} \left( x+3 \right)
 \left( x+25968 \right) \left( x+15827 \right) {x}^{17} \nonumber \\
 \qquad {}\times
\left( {x}^{6}+704 {x}^{5}+19667 {x}^{4}
+16573 {x}^{3}+221 {x}^{2}+5237 x+24649 \right)
 \nonumber \\
 \qquad {}\times P_{7}   P_{8}   P_{43}
 P_{22}  P_{16}  P_{13}
  {{Dx}}^{26} \nonumber \\
\qquad{}+3485 \left( x+10234 \right)^{11}   P_{109} {x}^{16}  \left( {x}^{5}+6389 {x}^{4}+9765 {x}^{3}
+14807 {x}^{2}+31264 x+20696 \right)
\nonumber \\
\qquad{}\times
 \left( {x}^{4}+24125 {x}^{3}+31305 {x}^{2}+16748 x+12080 \right)
 \left( x+30475 \right)   {{Dx}}^{25} \nonumber \\
\qquad{}+30663 \left( x+10234 \right)^{10}{x}^{15}
\left( {x}^{2}+4182 x+14901 \right)   P_{9}
  P_{49}  P_{60}  {{Dx}}^{24}
\nonumber\\
\qquad{}+286 \left( x+10234 \right)^{9} {x}^{14} \left( {x}^{3}+15593 {x}^{2}
+11835 x+28482 \right)   P_{100} \nonumber \\
\qquad {}\times \left( {x}^{3}+3549 {x}^{2}+28115 x
+25784 \right) \left( {x}^{3}+26209 {x}^{2}+12548 x+5267 \right)
   P_{12}  {{Dx}}^{23}
\nonumber \\
\qquad{}+ 3836 \left( x+10234 \right)^{8}   P_{28}  P_{20}
P_{66}  P_{7}   \left( x+495 \right)
 {x}^{13}   {{Dx}}^{22}
\nonumber \\
\qquad{}+16272 \left( x+10234 \right)^{7} \left( x+22272 \right) {x}^{12}
 \left( {x}^{2}+10657 x+16936 \right) P_{28}  P_{76}  P_{16}
  {{Dx}}^{21}
\nonumber \\
\qquad{} +108 \left( x+10234 \right) ^{6}{x}^{11} \left( x+12405 \right)
 \left( x+17141 \right)  \left( x+16254 \right)  P_{74}  P_{47}
{{\it Dx}}^{20}
\nonumber \\
\qquad{}+29777 \left( x+10234 \right)^{5} {x}^{10}
  P_{23}  P_{15}
P_{11}  P_{60}
\nonumber \\
\qquad {}\times
\left( {x}^{6}+23995 {x}^{5}+4070 {x}^{4}
+27561 {x}^{3}+19739 {x}^{2}+3632 x+18638 \right) \nonumber \\
\qquad {}\times
\left( {x}^{2}+24371 x+11409 \right) \left( x+22618 \right)\nonumber \\
\qquad {}\times
 \left( {x}^{4}+6026 {x}^{3}+10330 {x}^{2}+10566 x+27129 \right)
 \nonumber \\
\qquad {}\times
 \left( {x}^{3}+1171 {x}^{2}+10654 x+2741 \right)   {{Dx}}^{19}
\nonumber \\
\qquad{}+6325 \left( x+10234 \right)^{4}
\left( x+20731 \right) {x}^{9}
P_{42}  P_{25}  P_{51}
\left( {x}^{2}+16210 x+27515 \right)\nonumber \\
\qquad {}\times
 \left( {x}^{5}+11144 {x}^{4}+6536 {x}^{3}+25134 {x}^{2}
+10963 x+29010 \right)  {{Dx}}^{18}
\nonumber \\
\qquad{}+11986 \left( x+10234 \right)^{3}
 \left( {x}^{2}+16634 x+15614 \right)
 P_{40}  P_{13}  P_{72} {x}^{8}  {{\it Dx}}^{17}
\nonumber \\
\qquad{}+16154 \left( x+10234 \right)^{2} \left( {x}^{3}
+28948 {x}^{2}+8126 x+18460 \right) {x}^{7} \nonumber \\
\qquad {}\times
 \left( x+12041 \right) \left( x+9774 \right)
 P_{48}  P_{44}  P_{31}  {{Dx}}^{16}
\nonumber \\
\qquad{}+ 5724 \left( x+10234 \right) {x}^{6}
 P_{25}  P_{99} \nonumber \\
\qquad {}\times
\left( {x}^{4}+5795 {x}^{3}+12069 {x}^{2}+26629 x+1320 \right)
 \left( x+17629 \right)   {{Dx}}^{15}
\nonumber \\
\qquad{}+9093   P_{10}   P_{39}  P_{34}  P_{45}
\left( {x}^{2}+24077 x+23664 \right) {x}^{5}   {{Dx}}^{14}
\nonumber \\
\qquad{}+ 7105 {x}^{4}   P_{105}  P_{22}
\left( {x}^{3}+1115 {x}^{2}+1326 x+29632 \right)   {{Dx}}^{13}
\nonumber \\
\qquad{}+15661
\left( {x}^{5}+25773 {x}^{4}+9200 {x}^{3}
+26470 {x}^{2}+25643 x+1121 \right)
 P_{117}  P_{8}   {x}^{3}   {{Dx}}^{12}
\nonumber \\
\qquad{}+15107 \left( {x}^{4}+14631 {x}^{3}+6554 {x}^{2}+7715 x+3048 \right)
\left( x+6577 \right) \nonumber \\
\qquad {}\times \left( {x}^{3}+31616 {x}^{2}
+26256 x+9612 \right)   {x}^{2} \nonumber \\
\qquad {}\times
 \left( {x}^{3}+4756 {x}^{2}+28396 x+28874 \right)
  P_{10}   P_{58}  P_{51}  {{Dx}}^{11}
\nonumber \\
\qquad{}+26871 \left( {x}^{2}+11848 x+20401 \right)
 P_{11}   P_{48}  P_{69}  x   {{Dx}}^{10}
\nonumber \\
\qquad{}+541  P_{8}   Q_{8}   P_{14}  P_{66}  P_{33}
\left( x+932 \right)   {{Dx}}^{9}
+4081 \left( {x}^{2}+6617 x+3717 \right)   P_{45} P_{24}
  P_{18}  P_{20} \nonumber \\
\qquad {}\times \left( {x}^{3}+671 {x}^{2}+11514 x+23683 \right)
 \left( {x}^{2}+18485 x+5460 \right) \nonumber \\
\qquad {}\times
\left( {x}^{6}+23400 {x}^{5}+12243 {x}^{4}+21913 {x}^{3}
+27012 {x}^{2}+17751 x+12915 \right) \nonumber \\
\qquad {}\times \left( {x}^{2}+21416 x+31601 \right)   P_{7}
  {{Dx}}^{8}
+7551  P_{9}  P_{30}  P_{11}  P_{71}
 P_{7}  {{Dx}}^{7}
\nonumber \\
\qquad{}+6649  P_{73}  P_{26}  P_{23}
 \left( {x}^{3}+4619 {x}^{2}+29249 x+15768 \right)
 \left( {x}^{2}+12299 x+30824 \right)   {{Dx}}^{6}
\nonumber \\
\qquad{}+2785 \left( x+16825 \right)
 \left( x+27878 \right)
   \left( x+15523 \right)   P_{7}   P_{9}
  P_{102} \nonumber \\
\qquad {}\times
\left( {x}^{5}+13600 {x}^{4}+24394 {x}^{3}+31753 {x}^{2}
+19488 x+21782 \right)
  {{Dx}}^{5}
\nonumber \\
\qquad {}+14219  P_{18}  P_{105} \left( {x}^{2}+19404 x+23792 \right)
  {{Dx}}^{4}
\nonumber\\
\qquad{}+28992  P_{117}
 \left( {x}^{3}+9172 {x}^{2}+23091 x+20852 \right)
\left( x+15969 \right) \nonumber \\
\qquad {}\times \left( {x}^{2}+31441 x+5617 \right)
 \left( x+32359 \right)  {{Dx}}^{3}
\nonumber \\
\qquad{}+23799  P_{7}  P_{102}
\left( {x}^{3}+7169 {x}^{2}+15714 x+24623 \right) \nonumber \\
\qquad {}\times
\left( {x}^{6}+21946 {x}^{5}+31546 {x}^{4}+16591 {x}^{3}
+19174 {x}^{2}+23949 x+23818 \right) \nonumber \\
 \qquad {}\times \left( x+10738 \right)
 \left( {x}^{4}+10798 {x}^{3}+20551 {x}^{2}+6303 x+7193 \right)
  {{Dx}}^{2}
\nonumber \\
\qquad{}+  3515 \left( x+16461 \right) \left( x+6204 \right)
 P_{15}  P_{10}  P_{84} \left( x+4637 \right)
\left( x+1197 \right)\nonumber \\
 \qquad {}\times
 \left( {x}^{6}+31051 {x}^{5}+11003 {x}^{4}+8211 {x}^{3}
+28599 {x}^{2}+20034 x+25604 \right) \nonumber \\
 \qquad {}\times
 \left( x+17071 \right) \left( {x}^{2}+26670 x+1134 \right)  {Dx}
+ 25380 \left( x+30713 \right)  P_{35}
  P_{15}   P_{33}  P_{26}
\nonumber \\
 \qquad {}\times
 \left( {x}^{5}+8267 {x}^{4}+15086 {x}^{3}
+11158 {x}^{2}+26216 x+31098 \right)
\nonumber \\
 \qquad {}\times
 \left( {x}^{6}+30193 {x}^{5}+28390 {x}^{4}
+17930 {x}^{3}+26696 {x}^{2}+7578 x+16219 \right),
\nonumber
\end{gather*}
where the $ P_{n}$ and $ Q_{n}$ a ``short'' notation
for polynomials of degree $ n$ (that may be dif\/ferent from one
order $ D_x^m$ to another).
 The factor $\left( x +32748 \right)$ in the head polynomial
(coef\/f\/icient of $ Dx^{30}$) is nothing but the factor $\left( x-1 \right)$
modulo the prime $ 32749$.

\section[Singularities in the linear ODE for $\Phi_H^{(7)}$ and
$\Phi_H^{(8)}$]{Singularities in the linear ODE for $\boldsymbol{\Phi_H^{(7)}}$ and
$\boldsymbol{\Phi_H^{(8)}}$}
\label{singphi7phi8}

For $\Phi_H^{(7)}$, we generated long series, unfortunately, insuf\/f\/icient
to obtain the corresponding linear ODE.
Actually, we have also generated very long series modulo a prime
(40000 coef\/f\/icients) and we have not been able to f\/ind a linear ODE when the order
of the ODE is less than 100.
However, by steadily increasing the order $q$ of the ODE and the degrees $n$
of the polynomials in front of the derivatives, one may recognize, in
f\/loating point form, the singularities of the linear ODE as the roots of the
polynomial in front of the higher derivative.
A root is considered as singularity of the still unknown linear ODE, when
as $ q$ and $ n$ increase, it persists with more stabilized digits.

Using $1250$ terms in the series for $ \Phi_H^{(7)}$, the following
singularities are recognized
\begin{gather*}
 \left( 1-4 w\right) \left( 1-5 w+6 {w}^{2}-{w}^{3} \right)
 \left( 1+2 w-8 {w}^{2}-8 {w}^{3} \right) \left( 1+4 w\right)
 \left( 1+w \right)   w \nonumber \\
\qquad{}\times \left( 1+2 w-{w}^{2}-{w}^{3} \right)
 \left( 1-3 w+{w}^{2} \right)
\left( 1 +2 w -4 {w}^{2}\right) \left(1-7 w+5 {w}^{2}-4 {w}^{3} \right) \nonumber \\
\qquad{}\times \left(1-3 w-10 {w}^{2}+35 {w}^{3}+5 {w}^{4}
-62 {w}^{5}+17 {w}^{6}+32 {w}^{7}-16 {w}^{8} \right) \nonumber \\
\qquad{}\times \left(1+8 w+15 {w}^{2}-21 {w}^{3}
-60 {w}^{4}+16 {w}^{5}+96 {w}^{6}+64 {w}^{7}
\right) \nonumber \\
\qquad{}\times \left(1-4 w-16 {w}^{2}-48 {w}^{3}+32 {w}^{4}-128 {w}^{5} \right)
 \left(1-10 w+35 {w}^{2}-51 {w}^{3}+21 {w}^{4}-4 {w}^{5} \right)
 \nonumber \\
\qquad{}\times \left(1+7 w+26 {w}^{2}+7 {w}^{3}+4 {w}^{4}\right)
 \left(1+8 w+20 {w}^{2}+15 {w}^{3}+4 {w}^{4} \right) \nonumber \\
\qquad{}\times \left( 1+12 w+54 {w}^{2}+112 {w}^{3}+105 {w}^{4}
+35 {w}^{5}+4 {w}^{6} \right) = 0. \nonumber
\end{gather*}

Note that we have not seen with the precision of these calculations
the occurrence of the singularities of the $ \Phi_{H}^{(3)}$.

With similar calculations using $2000$ terms for $\Phi_H^{(8)}$, the
following singularities are recognized
\begin{gather*}
 \left( 1-2 w \right) \left( 1+2 w \right) \left(1-2 {w}^{2}
 \right) \left( 1-4 w \right) \left( 1-4 w+2 {w}^{2} \right)
 \left( 1+4 w \right) \left(1+ 3 w \right)  w \nonumber \\
 \qquad{}\times\left( 1+4 w+2 {w}^{2} \right) \left( 1-8 {w}^{2} \right)
 \left(1- 3 w \right) \left( 1-w \right)
 \left( 1+w \right) \left(1- 5 w \right) \left( 1+2 {w}^{2} \right) \nonumber \\
\qquad{}\times \left(1-26 {w}^{2}+242 {w}^{4}-960 {w}^{6}
+1685 {w}^{8}-1138 {w}^{10}\right) \left(1-10 {w}^{2} +32 {w}^{4} \right)\nonumber \\
\qquad{}\times \left(1 -30 {w}^{2}+56 {w}^{4}-1312 {w}^{6} \right)
 \left(1 -6 w+10 {w}^{2} \right)
 \left(1 -6 w+8 {w}^{2}-4 {w}^{3} \right) \nonumber \\
\qquad{}\times\left( 1+5 w \right) \left(1 +6 w+10 {w}^{2} \right)
 \left(1 +6 w+8 {w}^{2}+4 {w}^{3} \right)
 = 0. \nonumber
\end{gather*}

Note that the stabilized digits in these singularities can be as low
as two digits.

\section{Selected values for Liouville theory and Potts models}
\label{ee}

New classes of critical statistical models where
 suggested~\cite{Gervais} by Gervais and Neveu
 from the construction of Liouville f\/ield theory. With
the $ Q$-state standard scalar Potts model
notations (see~(1.3) in~\cite{Gervais}), they introduced $ y$, such that
$ Q^{1/2} = 2 \cos(\pi y/2)$. Rational
 values of $ y$ correspond to
selected values of $ Q$ (Tutte--Beraha numbers
see Section 4 of~\cite{Rammal})
for which the standard scalar Potts model has rational critical
 exponents. At this step, and in order to make explicit the selected role
of these particular values, we can recall the
expression (see (3.3) in~\cite{Rammal2}) of the partition function per site
 of the $ Q$-state standard scalar Potts model
on the checkerboard lattice in terms of
 Eulerian products (see (3.5) in~\cite{Rammal2})
like (with the notations of~\cite{Rammal2}):
\begin{gather}
\label{euler}
F(u) = \prod_{n=1}^{\infty}
 {{ 1 -t^{4 n-1} u} \over {1 -t^{4 n+1} u} },
\qquad  \hbox{where} \qquad  Q^{1/2} = t + {{1} \over {t}.}
\end{gather}
This Eulerian product form made very clear
 the fact that the partition function can be seen
as some automorphic function with respect to
 an inf\/inite discrete group generated by the
inverse relation and the symmetries of square~\cite{Jaekel2,Jaekel}.
Such Eulerian product over an inf\/inite discrete group also made very clear
the fact that these singled-out values of $ Q$ actually
 correspond\footnote{Note that this $ Q$, corresponding to the number
of state of the Potts model, should not be confused with a~nome $ q$.
It was unfortunately denoted $ q$ in~\cite{Gervais}.} to $ N$-th root of unity
situation
\begin{gather}
\label{nthroot}
 t^N = 1, \qquad \hbox{where} \qquad
 Q^{1/2} = t + {{1} \over {t}}
\end{gather}
that occur in some many domains of theoretical
physics~\cite{Schwartz,Todorov} (dilogarithms, Kac determinant, \dots).
Do note that such situation generalizes, mutatis mutandis, to the Baxter model:
the partition function per site can actually be written
as an inf\/inite discrete product~\cite{Baxter,Baxter2,Baxter3}
over a group generated by the inverse relation and
 geometrical symmetries of lattice~\cite{Maillard},
expressions like (\ref{euler}) being replaced by
(with Baxter's notations~\cite{Baxter,Baxter2,Baxter3})
\begin{gather*}
G(z) =
 \prod_{m=0}^{\infty} \prod_{n=1}^{\infty}
 {{ 1 -q^m x^{4 n-1} z} \over {1 -q^m x^{4 n+1} z} },
\end{gather*}
where
\begin{gather*}
q = \exp(-\pi {\cal I'} /{\cal I}), \qquad
x = \exp(-\pi \lambda /2{\cal I}) , \qquad
z = \exp(-\pi v /2{\cal I}). \nonumber
\end{gather*}

Such an expression of the partition function per site of the Baxter model
as inf\/inite product can also be found
in~\cite{Varchenko} in terms of product and ratio
of theta and elliptic gamma functions.

In~\cite{Gervais} Gervais and Neveu underlined that
 they had built Liouville f\/ield theory
for other singled-out values of $ Q$ than $ N$-th root of unity
situations like (\ref{nthroot}), namely (see~(2.3) in~\cite{Gervais})
\begin{gather*}
Q = -4 \sinh^2(\pi/2 \sqrt{3}),
\qquad  \hbox{and} \qquad
Q = 4 \cosh^2(\pi)
\end{gather*}
meaning respectively, in term\footnote{One has
to be careful with the various notations
in the literature where, as far as nomes are concerned,
 one moves from $ q$ to $ q^2$.
In~\cite{jFunc} the nome $ \bar{q}$ corresponds to $ t^2$. Relation
(\ref{othervaluesint2}) reads $ \bar{q} $
$ = q^2 = e^{2 i \pi \tau}
 = - e^{-\pi \sqrt{3} } $.} of $ t$
 def\/ined in (\ref{euler}) or (\ref{nthroot})
\begin{gather}
\label{othervaluesintE}
t = e^{ i \pi (1+i \sqrt{3})/2}
 = i   e^{-\pi \sqrt{3}/2 },
 \end{gather}
and
\begin{gather}
\label{othervaluesint2E}
t = e^{ i \pi (1+i)} = -e^{-\pi}
\end{gather}
that is $ t^2 = - e^{-\pi \sqrt{3}}$ and
 $ t^2 = - e^{-2 \pi }$
respectively.
Actually the variable $ t$ in (\ref{euler}) or
in~\cite{Rammal2} is {\em exactly what is called the multiplicative
crossing in conformal
 theory}~\cite{multiplicative3,multiplicative2,multiplicative1,multiplicative}.
Conformal f\/ield theoreticians are keen on introducing modular
group structure for which the multiplicative
crossing is seen as a modular nome $ q$. If we follow this
line recalling the relation $ q = \exp(i \pi \tau)$
 between the nome and the half period ratio $ \tau$
(see~\cite{jFunc}), we f\/ind that the two previous
situations actually correspond to
singled-out values of the modular $ j$-function namely
$ j((1+i \sqrt{3})/2) = (0)^3$ for (\ref{othervaluesintE}),
and $ j(1+i) = (12)^3$ for (\ref{othervaluesint2E}), which
actually correspond to {\em Heegner numbers
and complex multiplication}~\cite{broglie,jFunc}.

Considering $ \lambda$-extensions of two-point diagonal correlation
function of the Ising model, we found~\cite{Holo}
 modular curves corresponding to
polynomial relations between a (modular) function
and its f\/irst derivative, this (modular) function
being a {\em very simple ratio of Jacobi
theta functions} (see Section~6.1 in~\cite{Holo}). Along this line
 it is worth recalling the ``special value''
$ - e^{-\pi \sqrt{3}}$ of
the nome of Jacobi theta
 functions (at zero argument) for which
a ratio of Jacobi
theta functions becomes a simple algebraic expression~\cite{Borwein2}
\begin{gather*}
 {{ \theta_2(0, - e^{-\pi \sqrt{3} }) }
 \over {\theta_3(0, - e^{-\pi \sqrt{3} } )}}
 =
 \bigl( 4 \sqrt{3} -7 \bigr)^{1/4}.
\end{gather*}

At this step it is fundamental to raise an important
confusion that overwhelms the theoretical physics
 literature. In many domains of
theoretical physics the existence of a modular group and/or
$ N$-th root of unity situations in some ``nome'' always denoted
$ q$, is underlined and analyzed. In Liouville theory this nome $ q$
is the exponential\footnote{Not to be confused
 with the $ q$ of the $ q$-state
Potts model in the paper that cope with Liouville theory and Potts model
in the same time!} of $\hbar$, in conformal
 f\/ield theory\footnote{They are, of course, many other occurrences of modular groups
and/or occurrences of a nome $ q$ (quantum dilogarithms,
 $q$-deformation theories, $q$-dif\/ference
 equations, $q$-Painlev\'e, $q$-analogues
of hypergeometric functions, \dots). The confusion is
 increased with the dilute $ A_L$ models
and their relations with the Ising Model in a Field
for which the corresponding nome $ q$ could be
associated with the magnetic f\/ield of the
Ising model~\cite{E8bis,Seaton,E8ter,E8}.}
two $q$'s, and two modular group
structures, can be introduced, the second one corresponding
to f\/inite size analysis with the introduction
of a {\em modular parameter} for
partition function
on a~(f\/inite size $ l \times l'$) torus (see
for instance~(3.33) in~\cite{Pearce}).

Sticking with Baxter's notations the complex multiplication situation, we see in this paper with
selected values like $ 1 +3 w + 4 w^2 = 0$,
corresponds to selected values of the modulus of the elliptic curves,
or of the nome $ q$ which measures the distance to criticality (temperature-like variable) of the
 {\em off-critical lattice model}. In contrast the selected values (\ref{closed}) of $ \lambda$
(for which modular curves are seen to
occur for the $ \lambda$-extensions of the correlation functions)
correspond to $ N$-th root of unity situations for
the multiplicative crossing $ x$.
Most of the f\/ield theory papers (QFT, CFT, \dots) where
 selected values ($N$-th root of unity situations)
occur correspond to
models {\em at criticality}: for these models there is no
(temperature-like of\/f-critical) variable
like our previous nome $ q$ (the elliptic curve
is gone, being replaced by a rational curve). All the selected situations
encountered are in the multiplicative crossing variable $ x$
 within a rational parametrization of the model.

\section[Factorisations of multiple integrals linked to $ \zeta(3)$]{Factorisations of multiple integrals linked to $\boldsymbol{\zeta(3)}$}
\label{fin}

From the series expansion of the triple integral (\ref{otherzeta3})
we have obtained the corresponding order four Fuchsian
linear dif\/ferential equation ($Dx$ denotes $d/dx$)
\begin{gather*}
L_n = Dx^4
+\frac{2 (3 x -1)}{(x-1) x}   Dx^3
+\frac{\left(7 x^2 +(n^2+n-5) x
-2 n (n+1)\right)}{(x-1)^2 x^2 }   Dx^2
\nonumber \\
\phantom{L_n =}{}+ \frac{\left(x^2 +2 n (n+1)\right)}
{ (x-1)^2 x^3 }   Dx +\frac{n (n+1)
\left((n^2+n+1) x +(n-1) (n+2)\right)}{(x-1)^2 x^4}
\nonumber
\end{gather*}
which has the following factorization in order-one dif\/ferential operator:
\begin{gather*}
 L_n =
\left(Dx + {{d \ln(A_1)} \over {dx}} \right)
  \left(Dx + {{d \ln(A_2)} \over {dx}} \right)
  \left(Dx + {{d \ln(A_3)} \over {dx}} \right)
  \left(Dx + {{d \ln(A_4)} \over {dx}} \right), \nonumber
\end{gather*}
where the order-one dif\/ferential operators {\em have rational solutions} since:
\begin{gather*}
A_1 = -(n-1)  \ln(x) +2   \ln(x-1)
 +\ln(P_n), \nonumber \\
A_2 = (n+1)  \ln(x) -(n-1)   \ln(x-1)
 -\ln(P_n) +\ln(Q_n), \nonumber \\
A_3 = -n   \ln(x) +(n+1)   \ln(x-1)
 +\ln(P_n) -\ln(Q_n), \nonumber \\
A_4 = n   \ln(x) -\ln(P_n), \nonumber
\end{gather*}
and where $ P_n$ and $ Q_n$ are (normalized) polynomials in $ x$ of degree $ n$,
which satisfy, together with $P_n^{(m)}$ and $Q_n^{(m)}$ ($m = 1, \dots, 4$), their
 $ m$-th derivative
with respect to $ x$, a system of coupled dif\/ferential equations~\cite{bo-ha-ma-ze-07b}.

Such factorization in {\em order-one differential operator having rational solutions}
is characteristic of the {\em strong geometrical
 interpretation} we are seeking for (interpretation
of $ n$-fold integrals as periods of some algebraic variety)
for the Fuchsian linear dif\/ferential operators we have obtained for many
$ n$-fold integrals (of the ``Ising class''~\cite{crandall}).
Such a factorization in order-one linear dif\/ferential operator having rational solutions
does not seem to take place in general for our Fuchsian linear dif\/ferential operators,
{\em but seems actually to occur modulo many primes}
 for the Fuchsian linear dif\/ferential operators of the
$ \chi^{(n)}$. Such calculations, mixing
geometrical interpretation and ``modular'' calculations
on our $ n$-fold integrals, remain to be done.

\subsection*{Acknowledgements} We have deserved
 great benef\/it from discussions on various
aspects of this work with F.~Chyzak, G.~Delf\/ino, S.~Fischler, P.~Flajolet,
A.J.~Guttmann, M.~Harris, I.~Jensen, L.~Merel, G.~Mussardo,
B.~Nickel, J.H.H.~Perk, B.~Salvy, C.A.~Tracy and N.~Witte.
 We thank A.~Bostan for a search of linear
ODEs modulo primes with one of his magma program. We thank one of the three
referees for very usefull comments.
We acknowledge a CNRS/PICS f\/inancial support.
 One of us (NZ) would like to acknowledge kind hospitality
at the LPTMC where part of this work has been completed.
One of us (JMM) thanks the MASCOS (Melbourne) where
 part of this work was performed.

\pdfbookmark[1]{References}{ref}
\LastPageEnding

\end{document}